\documentclass{jfm}

\usepackage[utf8]{inputenc}
\usepackage[T1]{fontenc}
\usepackage{lmodern}

\usepackage{amsmath}
\usepackage{graphicx}
\usepackage{natbib}
\usepackage[hidelinks]{hyperref}
\usepackage{cleveref}
\usepackage{siunitx}   
\usepackage{microtype}

\newcommand*\lr[3]{{\left#1 #3 \right#2}}

\newcommand*\delim[4][\empty]{\mathopen{#1#2} #4 \mathclose{#1#3}}

\newcommand*\surf{\mathrm{s}}

\newcommand*\permut\varepsilon
\newcommand*\kd\delta
\newcommand*\dirac\delta
\newcommand*\Hstep\Theta
\DeclareMathOperator\Ind{\mathbb I}   

\newcommand*\dd{\mathop{}\!\mathrm{d}}
\newcommand*\pd{\partial}

\renewcommand*\vec{\boldsymbol}      
\newcommand*\uvec[1]{\hat{\vec#1}}  
\newcommand*\bvec{\hat{\vec\imath}}  
\newcommand*\ten[1]{\boldsymbol{\mathsfbi{#1}}}
\newcommand*\tsy{\mathsfi}
\newcommand*\vdot{\boldsymbol\cdot}  
\newcommand*\T{\mathsf T}           
\newcommand*\grad{\vec\nabla}
\newcommand*\gradWRT[1]{{\vec\nabla}_{\!\vec{#1}}}
\renewcommand*\div[1]{\grad\vdot#1}
\newcommand*\gradS{\grad_{\!\surf}}
\newcommand*\divS[1]{\gradS\vdot#1}
\newcommand*\laplace{\nabla^2}
\newcommand*\laplaceS{\nabla_{\!\surf}^2}
\newcommand*\oprod{\vec\otimes}
\newcommand*\Irr[1]{\left\{ #1 \right\}_0}

\newcommand*\FT{\widehat}

\DeclareMathOperator{\ExpInt}{E}

\newcommand*\scaled[1]{\tilde{#1}}   
\newcommand\Reals{\mathbb R}
\newcommand*\diracR[1]{\dirac_{\Reals^{#1}}}  
\newcommand*\para{\shortparallel}

\newcommand*\numCa{\mathinner{Ca}}  
\newcommand*\numBo{\mathinner{Bo}}  
\newcommand*\numMa{\mathinner{Ma}}  
\newcommand*\numBq{\mathinner{Bq}}  
\newcommand*\numPe{\mathinner{Pe}}  
\newcommand*\lenBq{L_\text{B}}      

\newcommand*\ga\alpha%
\newcommand*\gb\beta%
\newcommand*\gc\gamma%
\newcommand*\gd\delta%

\newcommand*\species{\textit}

\crefname{equation}{}{}  
\Crefname{equation}{Equation}{Equations}
\crefname{figure}{figure}{figures}  
\Crefname{figure}{Figure}{Figures}

\shorttitle{Driven and active colloids at fluid interfaces}
\shortauthor{N. G. Chisholm and K. J. Stebe}

\title{Driven and active colloids at fluid interfaces}
\author{%
  Nicholas G. Chisholm\aff{1}
  \and Kathleen J. Stebe\aff{1}
    \corresp{\email{kstebe@seas.upenn.edu}},
}
\affiliation{%
  \aff{1}Department of Chemical and Biomolecular Engineering,
         University of Pennsylvania, Philadelphia, PA 19104, USA
}

\begin{document}

\maketitle

\begin{abstract}
  We derive expressions for the leading-order far-field flows generated by externally driven and active (swimming) colloids at planar fluid--fluid interfaces.
  We consider colloids adjacent to the interface or adhered to the interface with a pinned contact line.
  The Reynolds and capillary numbers are assumed much less than unity, in line with typical micron-scale colloids involving air-- or alkane--aqueous interfaces.
  For driven colloids, the leading-order flow is given by the point-force (and/or torque) response of this system.
  For active colloids, the force-dipole (stresslet) response occurs at leading order.
  At clean (surfactant-free) interfaces, these hydrodynamic modes are essentially a restricted set of the usual Stokes multipoles in a bulk fluid.
  To leading order, driven colloids exert Stokeslets parallel to the interface, while active colloids drive differently oriented stresslets depending on the colloid's orientation.
  We then consider how these modes are altered by the presence of an incompressible interface, a typical circumstance for colloidal systems at small capillary numbers in the presence of surfactant.
  The leading-order modes for driven and active colloids are restructured dramatically.
  For driven colloids, interfacial incompressibility substantially weakens the far-field flow normal to the interface; the point-force response drives flow only parallel to the interface.
  However, Marangoni stresses induce a new dipolar mode, which lacks an analogue on a clean interface.
  Surface-viscous stresses, if present, potentially generate very long-ranged flow on the interface and the surrounding fluids.
  Our results have important implications for colloid assembly and advective mass transport enhancement near fluid boundaries.
\end{abstract}

\section{Intoduction}%
\label{sec:intro}

Fluid--fluid interfaces provide a rich setting for driven and active colloidal systems.
Here, a ‘driven’ colloid moves through a fluid due to external forces or torques, for example, a magnetic bead forced by a magnetic field.
‘Active’ colloids, on the other hand, self-propel by consuming a fuel source.
For example, motile bacteria are active colloids that self-propel by the rotation of one or more flagella.
Autophoretic nanorods or Janus particles are other examples of commonly studied active colloids.
These catalytic swimmers self-propel via generation of chemical gradients that produce a propulsive layer of apparent fluid slip along the colloid surface.

Past work on colloids adhered to interfaces has focused on their usefulness as Brownian rheological probes embedded in biological lipid membranes or surfactant monolayers, where colloid motion is, in this case, ‘driven’ by thermal fluctuations.
For example, colloidal probes have been used to measure surface viscosity of a fluid interface as a function of surfactant concentration \citep{Sickert2007}.
Such measurements require theoretical models of the mobility of the colloid.
\citet{Saffman1975} analytically computed the mobility of a flat disk embedded in a viscous, incompressible membrane separating two semi-infinite subphases in the limit of large Boussinesq number, a dimensionless number comparing the membrane viscosity to that of the surrounding fluid.
This calculation was extended to moderate Boussinesq numbers by \citet{Hughes1981} and to subphases of finite depth by \citet{Stone1998}.
Later theoretical work quantified the response of a linearly viscoelastic membrane to an embedded point force \citep{Levine2002}.
The effects of particle anisotropy have been quantified in the context of the mobility of a needle embedded in an incompressible Langmuir monolayer overlying a fluid of varying depth \citep{Fischer2004}.
Finally, the impact of interfacial compressibility and surfactant solubility on the drag on a disk embedded in an interface above a thin film of fluid has been quantified \citep{Elfring2016}.
The dynamics of (three-dimensional) colloids that protrude into the surrounding fluid phases has also been characterized.
Analytical and numerical analyses of the mobility of spheres \citep{Fischer2006,Pozrikidis2007,Stone2015,Doerr2015,Dani2015,Doerr2016} and thin filaments \citep{Fischer2006} can be found in the literature for clean and surfactant-laden interfaces in the limit of small capillary number, a dimensionless ratio of characteristic viscous stresses to interfacial tension.

Active colloids are also strongly influenced by fluid interfaces.
Motile bacteria have been extensively studied as biological active colloids due to their relevance to human health and the environment.
Seminal work by \citet{Lauga2006} showed, via a resistive-force theory model, that circular trajectories of \species{E.~coli} swimming near a solid boundary are caused by hydrodynamic interaction with the boundary.
Similar results are found for free surfaces \citep{DiLeonardo2011}, although the direction of circling is reversed.
The bacterium is also drawn toward the boundary by these hydrodynamic interactions.
More detailed boundary element simulations have shown the existence of stable trajectories of bacteria near solid boundaries, where the distance from the boundary and curvature of the trajectory reach a steady state \citep{Giacche2010}.
Thus, hydrodynamic interactions are one mechanism whereby bacteria may remain motile yet become trapped at the boundary.
In contrast, similar calculations show only unstable trajectories for swimmers near free surfaces; the swimmer inevitably crashes into the boundary unless it is initially angled steeply enough away to escape it altogether \citep{Pimponi2016}.
Finally, \citet{Shaik2017} analytically computed the motion of a spherical ‘squirmer,’ a common model for microorganism locomotion, near a weakly deformable interface.
Others have investigated the motion of autophoretic swimmers at fluid interfaces.
Gold--platinum catalytic nanorods are highly motile at alkane--aqueous
Further experiments have shown that partially wetted, self-propelled Janus particles at air--water interfaces move along circular trajectories with markedly decreased rotational diffusion as compared to their motion in a bulk fluid \citep{Wang2017}.
Theoretical analysis has yielded analytical predictions of the linear and angular velocities of an autophoretic sphere straddling a surfactant-free interface with a freely-slipping, \SI{90}{\degree} contact line \citep{Malgaretti2016}.
This work has supplied valuable information about the influence of fluid interfaces on active colloid locomotion.

Rather than developing detailed models for specific types of swimmers, an alternative approach is to use far-field models that capture universal features of colloid locomotion.
For active colloids, this approach has been used to compute swimming trajectories near solid boundaries \citep{Spagnolie2012} and fluid interfaces \cite{Lopez2014}.
Such methods are accurate when the colloid is separated from the boundary by a few body lengths \citep{Spagnolie2012}.
Recent work has employed far-field models of active colloids to study trapping of microswimmers near surfactant-laden droplets \citep{Desai2018} and the density distribution of bacteria near fluid interfaces \citep{Ahmadzadegan2019}.

Prior theoretical analyses have largely focused on computing drag on driven colloids or swimming trajectories of active colloids and how they are influenced by the boundary.
The actual flows generated by such colloids at interfaces and the implications of these flows have received less attention.
However, it is important to understand such flows, as they are of primary importance to interactions between colloids at the interface as well as enhanced mixing driven by colloid motion.

While trapping due to hydrodynamic interactions is well appreciated, there is another mechanism, unique to fluid interfaces, which can strongly alter the mobility and induced flows of driven or active colloids.
Fluid interfaces trap particles by their contact lines, where the fluid interface intersects the surface of the particle.
Such contact lines are ‘pinned,’ as they are essentially fixed relative to the particle's surface.
The wetting configurations on the particles relax very slowly, consistent with kinetically controlled changes in the location of the contact line \citep{Kaz2012,Colosqui2013}.
Detailed studies have documented contact-line pinning at asperities or high-energy sites on the surfaces of micron-scale polymeric particles \citep{Kaz2012,Wang2017}.
Because of the random nature of contact-line pinning, particles of a single type have a wide range of wetting configurations at the interface.
Recent research suggests that naturally occurring active colloids can also have pinned contact lines.
For instance, the bacterium \species{Pseudomonas aeruginosa} has been observed in a variety of different orientations at hexadecane--aqueous interfaces that persist over long times for each individual.
These different orientations of the body with respect to the interface are associated with distinct motility patterns \citep{Deng2020}.
More complex biohybrid colloids of \species{P. aeruginosa} adhered to polystyrene microbeads also exhibit a wide range of persistent, complex motions at fluid interfaces \citep{Vaccari2018}.
On interfaces with surface tensions typical for alkane--aqueous systems, like those considered here, contact-line pinning significantly constrains the motion of driven and active colloids.
Furthermore, we expect the fluid flow induced by driven or active colloids to be strongly influenced by their configuration relative to the interface.
Pinned contact lines allow particles to translate in the plane of the interface and rotate about the interface normal.
However, translation normal to the interface and rotation about an axis in the interface are precluded.
The hydrodynamic implications of such trapped states have not been discussed.

In this article, we use the multipole expansion method to derive the hydrodynamic modes generated by driven and active colloids at fluid interfaces.
We focus on the leading-order multipoles, which are expected to dominate the far-field flow and therefore may be observable in experiment.
We focus on the case where the colloid is physically adhered to a fluid interface with a pinned contact line that constrains its motion.
We also consider the case where the colloid is adjacent to the interface, as might occur due to hydrodynamic trapping.
By ‘adjacent,’ we mean that the colloid is wholly immersed in one of the fluids and is near but not touching the interface.
This article is organized as follows.
In \cref{sec:governing-eqs}, we develop the governing equations for the fluid motion due to colloids at two types of fluid interfaces: a clean, surfactant-free interface and an interface that is rendered incompressible by adsorbed surfactant.
In \cref{sec:reciprocal-relations}, we develop a reciprocal relation that applies to two fluids in Stokes flow separated by either of these types of interface.
In \cref{sec:clean-interfaces}, we develop a multipole expansion appropriate for colloids trapped at a clean interface, and we discuss the leading-order modes that are produced in the driven an active cases.
We then compare these results to analogous results at an incompressible interface in \cref{sec:incompressible-interfaces}.
Finally, we conclude in \cref{sec:conclusion} by discussing the implications of our results and opportunities for future research.

\section{Governing equations}
\label{sec:governing-eqs}

\subsection{Equations of motion}

\begin{figure}
  \centering
  \includegraphics[width=\linewidth]{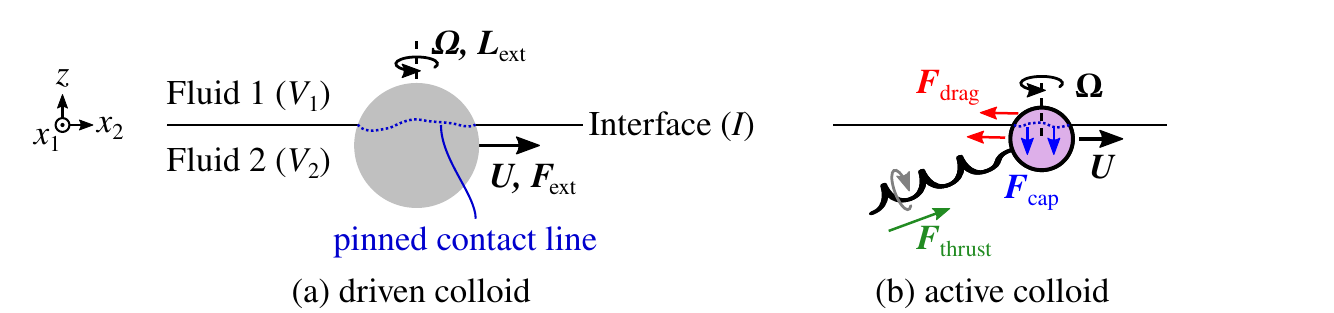}
  \caption{%
    Driven and active colloids at interfaces. Panel (a) is a colloid adhered to a fluid interface by a pinned contact line driven into motion by external force and torque fields.
    In response to this forcing, the colloid translates parallel to the interface at velocity $\vec U$ and rotates on the axis normal to the interface at angular velocity $\vec\Omega$.
    Other motions are prohibited due to contact-line pinning and elevated surface tension.
    Panel (b) is a similar illustration of an active colloid; we take a motile bacterium as a natural example.
    Thrust generated by the rotating helical flagellum is balanced by drag due to viscous dissipation and capillary forces.
  }%
  \label{fig:setup}
\end{figure}

We consider a colloid adhered to a planar interface between two immiscible  Newtonian fluids of viscosities $\mu_1$ and $\mu_2$, which are quiescent in the far field and together form an unbounded domain, as illustrated in \cref{fig:setup}.
As described in \cref{sec:intro}, we assume the resulting three-phase contact line is pinned, that is, it cannot move relative to the surface of the colloid.
For simplicity, we further assume that the interface is flat.
This assumption physically requires that (i) viscous stress due to flows generated by the particle are negligible compared to surface tension $\gamma$, which determines the equilibrium shape of the interface; (ii) the weight $mg$ of the colloid is also negligible compared to surface tension; and (iii) the amplitude of the undulations in the contact line are negligibly small compared to the size of the colloid.
Requirement (i) is formally satisfied when $\numCa = \mu U / \gamma \ll 1$, where $\numCa$ is the capillary number, $\mu$ is the fluid viscosity and $U$ is the characteristic velocity of the colloid.
For typical colloidal systems at air--aqueous or alkane--aqueous interfaces, $\numCa = O(\num{e-7})$ to $O(\num{e-5})$.
Requirement (ii) is satisfied when $\numBo = mg a^2 / \gamma \ll 1$, where $\numBo$ is the particle Bond number and $a$ is the characteristic length scale of the colloid.
In general, requirement (iii) may not be satisfied.
For isolated passive particles, nanometric contact-line distortions alter the capillary energy that traps colloids on interfaces \citep{Stamou2000},
and thermally activated fluctuations at the contact line are hypothesized to alter dissipation in the interface \citep{Boniello2015}.
Neither effect is included here, but the results we present may form the basis for a perturbative method to treat the problem of undulated contact lines.

At the colloidal scale, we may neglect the effects of fluid inertia and assume the flow on either side of the interface is governed by the Stokes equations,
\begin{equation}
  \label{eq:Stokes}
  \div{\ten\sigma}
  = - \grad p + \mu \laplace \vec u = \vec 0;
    \quad \div{\vec u} = 0,
\end{equation}
where \(\ten\sigma\) is the stress tensor, $\vec u$ is the fluid velocity, $p$ is the hydrodynamic pressure and \(\grad\) is the gradient operator.
The stress tensor is given by \(\ten\sigma = -p \ten I + \mu [ \grad\vec u + {(\grad\vec u)}^\T ]\), where $\ten I$ is the identity tensor.
These quantities vary with the position vector \(\vec x = x_1 \bvec_1 + x_2 \bvec_2 + z \bvec_3\).
Let $V_1$, $V_2$, and $I$ denote the set of points in fluid 1, fluid 2, and on the interface, respectively.
We assume that the viscosity changes abruptly across the interface as \(\mu(z) = \mu_1 \Ind_{\Reals_+}(z) + \mu_2 \Ind_{\Reals_-}(z)\), where the indicator function \(\Ind_P\) is unity if its argument is an element of \(P\) but otherwise vanishes (e.g., $\Ind_{\Reals_+}$ is equivalent to the Heaviside step function).
On the interface, \cref{eq:Stokes} satisfies the boundary conditions
\begin{subequations}
  \label{eq:BCs-on-I-general}
  \begin{align}
    \label{eq:BC-on-I--vel-continuity}
    \lr[]{\vec u}_I &= \vec 0 \\
    \label{eq:BC-on-I--no-penetration}
    \vec n \vdot \vec u(\vec x \in I) &= \vec 0 \\
    \label{eq:BC-on-I--stress-balance}
    \divS{\vec\varsigma} + \vec n \vdot \lr[]{\ten\sigma}_I \vdot \ten I_\surf &= \vec 0,
  \end{align}
\end{subequations}
where $\vec n$ is the unit normal to the interface pointing into fluid 1 and \(\lr[]{f}_I (\vec x \in I) := (\lim_{z \to 0^+} - \lim_{z \to 0^-}) f(\vec x)\) denotes the ‘jump’ in some function \(f = f(\vec x)\) across the interface going from fluid 2 to fluid 1.
The first two conditions assert that the fluid velocity is continuous across $I$ \cref{eq:BC-on-I--vel-continuity} and that fluid does not pass through the interface \cref{eq:BC-on-I--no-penetration}.
The last condition \cref{eq:BC-on-I--stress-balance} balances tangential stresses.
Here, \(\vec\varsigma = \vec\varsigma(\vec x \in I)\) is the surface stress tensor, \(\ten I_\surf = \ten I - \vec{nn}\) is the surface projection tensor, and \(\gradS = \ten I_\surf \vdot \grad\) is the surface gradient operator.
Note that, since we assume that the interface is planar, $\vec n = \bvec_3$ and $I$ is simply the set of points on \(z=0\).
Finally, as \(|\vec x| \to \infty\), the fluid velocity and pressure gradient in either volume vanish, i.e., \(\vec u(\vec x) \to \vec 0\) and \(p(\vec x) \to p_\infty\).

\subsection{Clean interface}

We call an interface ‘clean’ if it is free of surfactant molecules.
In the absence of temperature gradients, a clean interface is characterized by a uniform surface tension \(\gamma_0\), and \(\vec\varsigma(\vec x) = \gamma_0 \ten I_\surf\).
Then, $\divS{\vec\varsigma}$ vanishes and \cref{eq:BC-on-I--stress-balance} reduces to
\begin{equation}
  \label{eq:iface-stress-bal-clean}
  \vec n \vdot \lr[]{\ten\sigma}_I \vdot \ten I_\surf = \vec 0,
\end{equation}
which states that the tangential stress on the fluid is continuous across the interface.

\subsection{Incompressible interface}
\label{sec:governing-eqs-incompressible}

If surfactant is present, gradients in surfactant concentration due to flow exert Marangoni stresses on the surrounding fluids.
At interfaces where $\numCa \ll 1$, these gradients need only be infinitesimal to balance viscous stresses due to colloid motion.
As a result, the interface is constrained to surface-incompressible motion.

To derive the most conservative estimate for the effects of these Marangoni stresses, consider trace surfactant concentrations, for which the surfactant can be approximated as a two-dimensional ideal gas.
We define the surface pressure as \(\pi(\vec x \in I) = \gamma_0 - \gamma(\vec x \in I)\).
In this case, the dependence of the surface pressure on surfactant concentration $\Gamma = \Gamma(\vec x \in I)$ is given by $\pd\pi / \pd\Gamma = k_B T$, where $k_B$ is Boltzmann's constant and $T$ is temperature.
Scaling the surface pressure by viscous stresses \(\scaled\pi = \pi / \bar\mu U\), where \(\bar\mu = (\mu_1 + \mu_2)/2\) is the average surface viscosity, and letting \(\scaled\Gamma = \Gamma / \bar\Gamma\), where $\bar\Gamma$ is the average surface concentration over the entire interface, we find that
\( \scaled\gradS \scaled\pi = \numMa\, \scaled\gradS \scaled\Gamma \),
where the dimensionless parameter \(\numMa = {k_B T \bar\Gamma}/{\bar\mu U}\) is the Marangoni number and $\scaled\gradS = a\gradS$.
To evaluate $\numMa$, we consider typical parameter values for a colloid moving at $U = \SI{10}{\micro\meter/\second}$ at a hexadecane-water interface (\(\gamma_0 \approx \SI{50}{\milli\newton/\meter}\)) in the surface-gaseous state.
The surfactant concentration required to produce a \SI{0.1}{\percent} decrease in the surface tension is approximately \(\bar\Gamma = \SI{2e3}{\text{molecules}/\micro\meter^2}\).
Given $\bar\mu \approx \SI{1}{\milli\pascal\second}$, we estimate that $\numMa = O(10^3)$.
Thus, very small perturbations in $\Gamma$ generate sufficient Marangoni stress to balance viscous stresses due to motion of the colloid.

The large-$\numMa$ limit has the important consequence that the fluid interface behaves as incompressible layer (\(\divS{\vec u} = 0\)).
Assuming bulk-insoluble surfactant, the non-dimensionalized surfactant mass balance on the interface is
\begin{equation}
  \label{eq:mass-transport-nd}
  \scaled\Gamma(\vec x) \, \scaled\gradS \vdot \scaled{\vec u}
  + \numMa^{-1} (\scaled{\vec u} \vdot \scaled\gradS) \scaled\pi
  = {(\numMa\,\numPe_\surf)}^{-1} \scaled\laplaceS \scaled\pi,
\end{equation}
where \(\scaled{\vec u} = \vec u / U\).
Here, \(\numPe_\surf = U a / D_\surf\) represents the ‘interfacial’ Péclet number, where $D_\surf$ is the surface diffusivity of the adsorbed surfactant.
\Cref{eq:mass-transport-nd} implies that \(\scaled\gradS \vdot \scaled{\vec u} \ll 1\) if \(\numMa \gg 1\) and \(\numPe_\surf \gtrsim \numMa^{-1}\).
Assuming \(a = \SI{10}{\micro\meter}\) and \(D_\surf = \SI{e2}{\micro\meter^2/\second}\) (a typical value for small molecule surfactants), we have $\numPe_\surf = O(1)$, so surfactant diffusion does not restore compressibility of the interface.
At larger surfactant concentrations, the interface, populated by bulk-insoluble surfactants, generally departs from the surface-gaseous state.
The interface generally remains incompressible in this case because, excluding phase transitions, \(\pd\gamma / \pd\Gamma > k_B T\).
Thus, we hereafter assume \(\gradS \vdot \vec u = 0\) while discussing interfaces with surfactant.
Dilute soluble surfactants also obey this constraint, as mass transport rates between the bulk and the interface are typically negligible.
Note that we may express the Marangoni number as \(\numMa = E / {\numCa}\), where \(E = -(\bar\Gamma / \gamma) (\pd\gamma / \pd\Gamma)\) is the Gibbs elasticity.
Thus, interfacial incompressibility is the typical circumstance for interfacial flow at low capillary number \citep{Blawzdziewicz1999}.

Surfactants can also create surface-viscous stresses due to shearing motion of the interface.
If we assume Newtonian behavior, the interfacial stress tensor is given by
\begin{equation}
  \label{eq:iface-stress-tensor-incompr}
  \vec\varsigma(\vec x) = -\pi(\vec x) \ten I_\surf + \mu_\surf \left[ \gradS\vec u + {(\gradS\vec u)}^\T \right]
\end{equation}
for \(\vec x \in I\), where $\mu_\surf$ is the surface viscosity.
Then, \cref{eq:iface-stress-tensor-incompr} and \cref{eq:BC-on-I--stress-balance} yield the tangential stress balance for an incompressible, surfactant-laden interface,
\begin{equation}
  \label{eq:iface-stress-bal-Newtonian}
  - \gradS\pi + \mu_\surf \laplaceS \vec u
  + \vec n \vdot \lr[]{\ten\sigma}_I \vdot \ten I_\surf = \vec 0.
\end{equation}
\Cref{eq:iface-stress-bal-Newtonian} together with the incompressibility condition, \(\divS{\vec u} = 0\), are the Stokes equations for a two-dimensional Newtonian fluid being externally forced by bulk-viscous stresses.

\section{Reciprocal relation for two fluids separated by an interface}%
\label{sec:reciprocal-relations}

\subsection{Lorentz reciprocal theorem across an interface}

\begin{figure}
  \centering
  \includegraphics{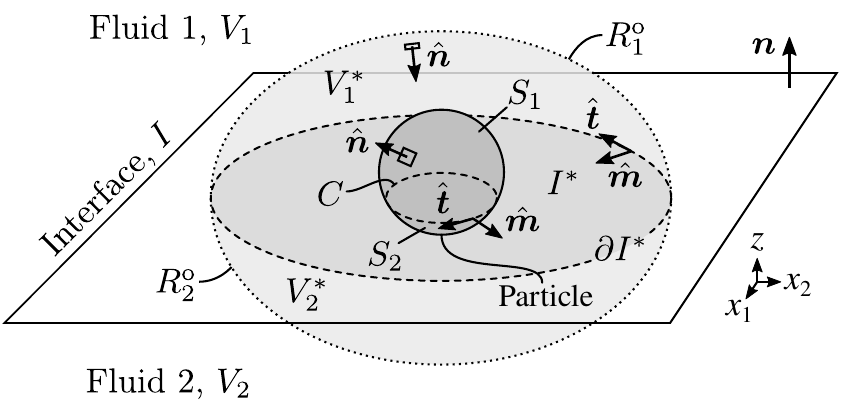}
  \caption{%
    A colloidal particle, depicted in the center of the illustration, is surrounded by two arbitrary fluid regions $V^*_1 \subset V_1$ and $V^*_2 \subset V_2$, which meet at region $I^* \subset I$ on the interface.
    We assign the inward-facing normal vector $\uvec n$ to the boundaries of both of these regions.
    The unit normal to the interface $\vec n$ (sans hat) points in the $+z$ direction.
    The boundary of $V_1^*$ consists of the colloid surface $S_1$, the interfacial region $I^*$, and the remaining outer surface $R^\text{o}_1$, with the boundaries of $V_2^*$ being similarly labeled.
    The boundary of $I^*$ (dashed line), denoted $\partial I^*$, has the counterclockwise-oriented tangent vector $\uvec t$, and we define \(\uvec m = \vec n \times \uvec t\), which points into $I^*$.
    The three-phase contact line $C$ comprises the inner part of $\partial I^*$.
  }%
  \label{fig:space}
\end{figure}

The Lorentz reciprocal theorem provides a relation between the velocity and stress fields of two arbitrary Stokes flows.
We may extend this theorem to two fluid regions separated by a clean or incompressible Newtonian interface as follows.
Consider a region $V_\nu^* \subset V_\nu$ that is fully contained in fluid $\nu$, where \(\nu = 1 \text{ or } 2\), as illustrated by \cref{fig:space}.
Let $(\vec u, \ten\sigma)$ and $(\vec u', \ten\sigma')$ represent the velocity and stress fields of two different solutions to the inhomogeneous Stokes equations,
\begin{equation}
  \label{eq:Stokes-inhomog}
  \div{\ten\sigma} = -\vec f(\vec x); \qquad
  \div{\ten\sigma'} = -\vec f'(\vec x),
\end{equation}
for \(\vec x \in V^*_\nu\), each subject to the conditions given by \cref{eq:BCs-on-I-general} at the interface.
Here, the forcing functions $\vec f$ and $\vec f'$ aid in our generalization of the reciprocal theorem, as is customary in such derivations \citep{Kim1991}.
We will later assert that these quantities vanish.
Integration of \(\div{(\ten\sigma \vdot \vec u' - \ten\sigma' \vdot \vec u)}\) over $V_\nu^*$ and application of the divergence theorem leads to the identity \citep[see, e.g.,][]{Kim1991}
\begin{equation}
  \label{eq:R-ident-bulk}
  \int_{V_\nu^*} [(\div{\ten\sigma}) \vdot \vec u' - (\div\ten\sigma') \vdot \vec u] \dd V
  + \int_{\partial V_\nu^*} (\ten\sigma \vdot \vec u' - \ten\sigma' \vdot \vec u) \vdot \dd{\vec S}
  = 0
\end{equation}
where $\partial V_\nu^*$ denotes the boundary of $V_\nu^*$, and $\dd{\vec S} = \uvec n \dd S$ points into $V_\nu^*$.
Substituting \cref{eq:Stokes-inhomog} into \cref{eq:R-ident-bulk} gives the Lorentz reciprocal theorem,
\begin{equation}
  \label{eq:R-thm-bulk}
  \int_{V_\nu^*} [\vec f(\vec x) \vdot \vec u' - \vec f'(\vec x) \vdot \vec u] \dd V
  = \int_{\partial V_\nu^*} (\ten\sigma \vdot \vec u' - \ten\sigma' \vdot \vec u) \vdot\dd{\vec S}.
\end{equation}

We add the pair of equations given by \cref{eq:R-thm-bulk} for each of the two fluid phases (\(\nu = 1, 2\)) to obtain
\begin{multline}
  \label{eq:R-thm-sandwich}
  \int_{V^*} [\vec f(\vec x) \vdot \vec u' - \vec f'(\vec x) \vdot \vec u] \dd V \\
  = \oint_R (\ten\sigma \vdot \vec u' - \ten\sigma' \vdot \vec u) \vdot\dd{\vec S}
  + \int_{I^*} \left( \lr[]{\ten\sigma}_I \vdot \vec u'- \lr[]{\ten\sigma'}_I \vdot \vec u \right) \vdot \vec n \dd A,
\end{multline}
where $V^* := V_1^* \cup V_2^*$ is the union of the fluid volumes in each phase, $I^* := \partial V_1 \cap \partial V_2$ is the region (at the fluid interface) where $V_1^*$ and $V_2^*$ ‘touch’, and $R := \partial V^* \setminus I^*$ constitutes the remaining boundaries of $V_1^*$ and $V_2^*$ that are not adjacent to each other.
For example, for the fluid region illustrated in \cref{fig:space}, \(R = S_1 \cup S_2 \cup R^\text{o}_1 \cup R^\text{o}_2\), which includes both the surfaces of the colloid (the inner surfaces of $V^*$) and the outer surfaces of $V^*$.
Note that $V^*_1$ and $V^*_2$ are disjoint subsets of $V^*$; they do not include points on $I$.
We interpret the integral over $V^*$ in \cref{eq:R-thm-sandwich} as being a sum of integrations over each of these subsets, and we similarly interpret the integral over $R$.
In the integral over $I^*$, we have used the fact that the fluid velocities $\vec u$ and $\vec u'$ are continuous across the interface \cref{eq:BC-on-I--vel-continuity}.
This term can be recast using the interfacial stress balance;
contracting an arbitrary vector $\vec t^*$ directed tangent to the interface with \cref{eq:BC-on-I--stress-balance} gives
\begin{equation}
  \label{eq:iface-tangential-stress-bal-forced}
  (\divS{\vec\varsigma}) \vdot \vec t^*
  + \vec n \vdot \lr[]{\ten\sigma}_I \vdot \vec t^*
  + \vec f_\surf \vdot \vec t^*
  = 0,
\end{equation}
where we have included an additional external surface force density $\vec f_\surf = \vec f_\surf(\vec x \in I)$ on the interface.
Since there is no fluid flux through interface, both $\vec u$ and $\vec u'$ are tangent to the interface for $\vec x \in I$.
Thus, \cref{eq:R-thm-sandwich} and \cref{eq:iface-tangential-stress-bal-forced} give, after replacing $\vec t^*$ with $\vec u$, 
\begin{multline}
  \label{eq:R-thm-general-iface}
  \int_{V^*} \left[
    \vec f \vdot \vec u' - \vec f' \vdot \vec u
  \right] \dd{V}
  + \int_{I^*} \left[
    \vec f_\surf \vdot \vec u' - \vec f_\surf' \vdot \vec u
  \right] \dd A \\
  = \oint_{R} \left(
    \ten\sigma \vdot \vec u' - \ten\sigma' \vdot \vec u
  \right) \vdot \dd{\vec S}
  - \int_{I^*} \left[
    (\divS{\vec\varsigma}) \vdot \vec u'
    - (\divS{\vec\varsigma'}) \vdot \vec u
  \right] \dd{A},
\end{multline}
where $\vec\varsigma$ and $\vec\varsigma'$ are the interfacial stress tensors associated with the unprimed and primed flows, respectively.

\subsection{Clean interface}

For a clean interface, $\vec\varsigma = -\ten I_\surf \gamma_0$ is constant, so the final integral in \cref{eq:R-thm-general-iface} vanishes;
\begin{equation}
  \label{eq:R-thm-clean-iface}
  \int_{V^*} \left[
    \vec f \vdot \vec u' - \vec f' \vdot \vec u
  \right] \dd{V} +
  \int_{I^*} \left[
    \vec f_\surf \vdot \vec u' - \vec f_\surf' \vdot \vec u
  \right] \dd{A}
  = \oint_{R} \left(
    \ten\sigma \vdot \vec u' - \ten\sigma' \vdot \vec u
  \right) \vdot \dd{\vec S}.
\end{equation}
If we set \(\vec f_\surf = \vec f_\surf' = \vec 0\), then the integral over $I^*$ in \cref{eq:R-thm-clean-iface} also vanishes, which is the same as \cref{eq:R-thm-bulk} with $\partial V_\nu^*$ replaced by $R$.

\subsection{Incompressible interface}

Assuming an incompressible interface with Newtonian behavior, as described by \cref{eq:iface-stress-tensor-incompr}, there is a ‘surface’ reciprocal identity for the interface analogous to \cref{eq:R-thm-bulk} given by
\begin{equation}
  \label{eq:R-id-surface}
  \int_{I^*} \left[
    (\divS{\vec\varsigma}) \vdot \vec u' - (\divS\vec\varsigma') \vdot \vec u
  \right] \dd A
  + \oint_{\partial I^*} {%
    (\vec\varsigma \vdot \vec u' - \vec\varsigma' \vdot \vec u) 
  } \vdot \uvec m \dd C
  = 0,
\end{equation}
where the contour integral is taken over the boundary of $I^*$, denoted $\partial I^*$.
For the system of a particle on an interface illustrated in \cref{fig:space}, $\partial I^*$ includes the contact line on the particle $C$ as its inner boundary.
We assign \(\partial I^*\) the unit tangent vector \(\uvec t\) regarding $I^*$ as a counterclockwise-oriented surface.
The unit vector $\uvec m = \vec n \times \uvec t$ points into the interfacial region $I^*$, meeting $\partial I^*$ at a right angle.
\Cref{eq:R-id-surface,eq:R-thm-general-iface} yield
\begin{multline}
  \label{eq:R-thm-incompressible-iface}
  \int_{V^*} \left[
    \vec f \vdot \vec u' - \vec f' \vdot \vec u
  \right] \dd{V}
  + \int_{I^*} \left[
    \vec f_\surf \vdot \vec u' - \vec f_\surf' \vdot \vec u
  \right] \dd A \\
  = \oint_{R} \left(
    \ten\sigma \vdot \vec u' - \ten\sigma' \vdot \vec u
  \right) \vdot \dd{\vec S}
  + \oint_{\partial I^*} \left(
    \vec\varsigma \vdot \vec u' - \vec\varsigma' \vdot \vec u
  \right) \vdot \uvec m \dd C.
\end{multline}
Comparing \cref{eq:R-thm-incompressible-iface} to the analogous equation for a clean interface \cref{eq:R-thm-clean-iface}, we see that the final integral on the right-hand side is new.
This contour integral over the boundary of $I^*$ accounts for surface pressure gradients, or Marangoni stresses, that enforce the interfacial incompressibility constraint and, if $\mu_\surf > 0$, for surface-viscous dissipation.
While we restrict ourselves to planar interfaces, \cref{eq:R-thm-clean-iface,eq:R-thm-incompressible-iface} hold even if the interface is curved, given that it has the same shape in both the primed and unprimed flow problems.

\section{Clean fluid interfaces}
\label{sec:clean-interfaces}

While incompressible interfaces are typical for colloidal systems, as described in \cref{sec:governing-eqs-incompressible}, it is instructive to first consider clean interfaces.
In this section, we develop the multipole expansion for a colloid at a clean interface.
In \cref{ssec:clean-Green}, we review the Green's function for a clean interface, originally developed by \citet{Aderogba1978}.
Then, in \cref{ssec:clean-BIE}, we use \cref{eq:R-thm-clean-iface} to derive a boundary integral representation of the velocity field appropriate for developing the multipole expansion, which is done in \cref{ssec:clean-MPX}.
Finally, we discuss the implications of the leading-order multipoles for driven and active colloids on interfaces.

\subsection{Green's function}
\label{ssec:clean-Green}

Due to the linearity of the Stokes equations \cref{eq:Stokes} and its boundary conditions for a clean interface \cref{eq:iface-stress-bal-clean}, we may represent the velocity field due to a point force located at \(\vec y = y_1 \bvec_1 + y_2 \bvec_2 + h \bvec_3\) as \(\vec u(\vec x) = \ten G(\vec x, \vec y) \vdot \vec F\),
where \(\ten G\) is the Green's function for two fluids separated by a clean interface.
This Green's function satisfies the (inhomogeneous) Stokes equations
\begin{subequations}
  \label{eq:Stokes-G}
  \begin{align}
    - \grad \vec P(\ten G; \vec x, \vec y)
    + \mu(z) \, \laplace \ten G(\vec x, \vec y) &=
    \begin{cases}
      \vec 0                                     &  h = 0    \\
      -\ten I \diracR3(\vec x - \vec y)          &  h \neq 0
    \end{cases}
    \\
    \label{eq:incompressible-G}
    \div{\ten G(\vec x, \vec y)} &= \vec 0
  \end{align}
\end{subequations}
for \(\vec x \in V_1 \cup V_2\).
\Cref{eq:Stokes-G} is subject to the far-field condition \(\ten G \to \vec 0\) as \(|\vec x| \to \infty\), the interfacial stress balance
\begin{subequations}
  \label{eq:BCs-on-G}
  \begin{align}
    \label{eq:iface-stress-bal-G}
    \ten I_\surf \vdot \lr[]{\vec n \vdot \ten T(\ten G; \vec x, \vec y)}_I &=
    \begin{cases}
      -\ten I_\surf \diracR2(\vec x - \vec y)  &  h = 0      \\
      \vec 0                               &  h \neq 0,
    \end{cases}
    \intertext{and the kinematic conditions}
    \label{eq:iface-kinematics-G}
    \vec t^* \vdot \lr[]{\ten G(\vec x, \vec y)}_I &=
    \vec n \vdot \ten G(\vec x \in I) = \vec 0.
  \end{align}
\end{subequations}
Here, \(\vec P(\ten G;\!)\) is the (vectorial) pressure field associated with \(\ten G\), \(\ten T(\ten G;\!)\) is the stress tensor associated with \(\ten G\), and $\diracR{n}(\vec x)$ is the Dirac delta in $\Reals^n$.
As expressed by \cref{eq:iface-stress-bal-G}, for $h = 0$, we consider the point force to be exerted on the interface itself rather than on one of the fluids.
\Cref{eq:Stokes-G,eq:BCs-on-G} follow directly from \cref{eq:Stokes,eq:BCs-on-I-general} after factoring out $\vec F$ from both sides of each of these equations.

\begin{figure}
  \centering
  \includegraphics[width=\linewidth]{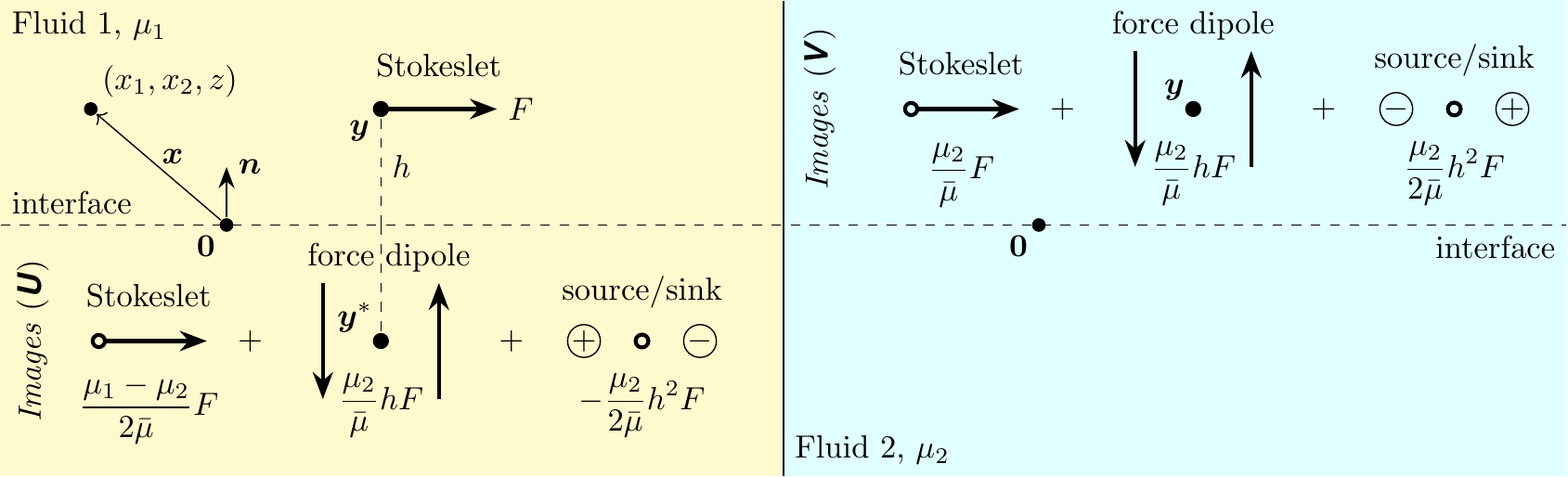}
  \caption{%
    A point force of magnitude \(F = |\vec F|\) parallel to the interface induces the image system illustrated above, as expressed by \cref{eq:Gfun-clean}.
    The upper left and lower right portions of the figure represent the physical fluid phases while the upper right and lower left are fictitious image domains that contain singularities acting to satisfy the boundary conditions on the interface \cref{eq:BCs-on-G}.
    The image singularities in each image domain are depicted separately for clarity, but they all act at the same image point, i.e., $\vec y^*$ for fluid 1 and $\vec y$ for fluid 2.
  }%
  \label{fig:clean-image-system}
\end{figure}


Solving \cref{eq:Stokes-G,eq:BCs-on-G} yields
\begin{equation}
  \label{eq:Gfun-clean}
  \ten G(\vec x, \vec y) =
  \begin{cases}
    [\ten J(\vec x - \vec y) + \ten U(\vec x, \vec y^*)] / \mu(h)
    & zh \geq 0 \\
    \ten V(\vec x, \vec y) / \bar\mu & zh \leq 0 \\
    \ten J(\vec x - \vec y) \vdot \ten I_\surf / \bar\mu & h = 0,
  \end{cases}
\end{equation}
where $\vec y^* = (y_1, y_2, -h)$ is the reflection of \(\vec y\) through \(z=0\).
\Cref{eq:Gfun-clean} expresses $\ten G$ as a functional of the Oseen tensor \(\ten J(\vec x) = (\ten I / |\vec x| - \vec x \vec x / |\vec x|^3) / 8\upi\).
The tensors $\ten U$ and $\ten V$ represent the hydrodynamic images necessary to satisfy continuity of tangential stress \cref{eq:iface-stress-bal-G} and continuity of velocity at the interface.
These image systems are given by \citep{Aderogba1978}
{\newcommand\yDn{\vec\xi \vdot \vec n}
\begin{align}
  \label{eq:upper-Blake-images}
  \tsy U_{ij}(\vec x, \vec\xi) &=
    (\kd^\para_{jk} - n_j n_k) \tsy J_{ik}(\vec x - \vec\xi) -
    \frac{\mu(\yDn)}{\bar\mu} \tsy V_{ik}(\vec x, \vec\xi) \\
  \label{eq:lower-Blake-images}
  \tsy V_{ij}(\vec x, \vec\xi) &= \left\{
    \kd^\para_{jk} +
    (\kd^\para_{jk} - n_j n_k) \left[
      (\yDn) n_l \frac{\pd}{\pd\xi_k} - \frac12 (\yDn)^2 \kd_{kl} \nabla^2
    \right]
  \right\} \tsy J_{il}(\vec x - \vec\xi),
\end{align}}%
where $\kd_{jk}$ is the Kronecker delta and $\kd^\para_{jk} = \kd_{jk} - n_j n_k$.
The tensor indices \(i,j,k,l \in \{1,2,3\}\) follow the Einstein summation convention.

\Cref{eq:Gfun-clean} partitions \(\ten G(\vec x, \vec y)\) into three cases: \(zh \geq 0\), where $\vec x$ and $\vec y$ are in the same fluid; \(zh \leq 0\), where $\vec x$ and $\vec y$ are in different fluids; and \(h = 0\) for all $z$, where \(\vec y \in I\).
Without loss of generality, if we let the point force at $\vec y$ be in the upper fluid ($h > 0$), then a Stokeslet, the fundamental solution to the Stokes equations in an unbounded fluid given by \(\ten J / 8 \upi \mu_1\), is induced at this point.
Examining, \cref{eq:lower-Blake-images}, the flow in the lower fluid ($z < 0$) comprises three image flows: a Stokeslet parallel to the interface, a Stokeslet dipole, and a degenerate Stokes quadrupole (a source--sink doublet), all of which have their singular points at $\vec y$.
These images are depicted on the upper right-hand (cyan) side of \cref{fig:clean-image-system}.
The image system $\ten U$ for the upper fluid \cref{eq:upper-Blake-images} is similar except that the image singularities are located at the image point $\vec y^*$, depicted on the lower left-hand (yellow) side of \cref{fig:clean-image-system}.
This image system additionally includes a copy of the original forcing Stokeslet reflected through $z=0$.

Finally, we note two properties of \(\ten G\) that will be useful in the analysis that follows.
First, it is self-adjoint,
\begin{equation}
  \label{eq:self-adjoint-G}
  \ten G(\vec x, \vec y) = \ten G^\T(\vec y, \vec x),
\end{equation}
which may be proven using \cref{eq:R-thm-clean-iface} (see \cref{sec:app:self-adjointness}) or directly verified from \cref{eq:Gfun-clean}.
The second property concerns the limiting behavior of $\ten G$.
As $|\vec x|$ becomes large for fixed $|\vec y|$, \(\ten G(\vec x, \vec y) \vdot \vec F\) effectively appears as a Stokeslet and decays as \({|\vec x - \vec y|}^{-1} \sim {|\vec x|}^{-1}\);
the image Stokes dipole and degenerate quadrupole terms, contained in \(\ten U\) and \(\ten V\), do not affect the far-field behavior of \(\ten G\) because their spatial decay is more rapid than that of the Stokeslet.
An exception occurs when $\vec F$ points directly away from the interface, in which case $\ten G \vdot \vec F$ reduces to an effective stresslet of strength \(h|\vec F| \mu(h) / \bar\mu\) for $|\vec x| \gg |\vec y|$ and decays as ${|\vec x|}^{-2}$ \citep{Aderogba1978}.
By \cref{eq:self-adjoint-G}, the decay behavior of $\ten G$ for fixed $\vec x$ as $\vec y$ is made large reflects the behavior for fixed $\vec y$ as $\vec x$ is made large; \(\ten G(\vec x, \vec y) \sim {|\vec y|}^{-1}\) for \(\vec y \gg \vec x\).

\subsection{Boundary integral equation}
\label{ssec:clean-BIE}

Using the Green's function \cref{eq:Gfun-clean} as the ‘primed’ flow field in the reciprocal relation \cref{eq:R-thm-clean-iface}, we may generate a boundary integral equation for an object at the interface.
Consider the interfacially trapped colloid, illustrated in \cref{fig:space}, whose upper surface \(S_1\) is in contact with fluid 1 and whose lower surface \(S_2\) is in contact with fluid 2.
An arbitrary volume of fluid \(V^* = V^*_1 \cup V^*_2\) surrounds the colloid, which is bounded by \(S_1\) and \(S_2\) as well as the outer fluid surfaces \(R^\text{o}_1\) and \(R^\text{o}_2\).
We make the following substitutions into \cref{eq:R-thm-clean-iface}:
\(\vec u'(\vec x) \to \ten G(\vec x, \vec y)\),
\(\ten\sigma'(\vec x) \to \ten T(\ten G; \vec x, \vec y)\),
\(\vec f' \to \ten I \diracR3(\vec x - \vec y)\), and
\(\vec f_\surf' \to \ten I \diracR2(\vec x - \vec y)\) to find
\begin{multline}
  \label{eq:BIE-simp-step-0}
  \int_{V*} \delim[\big][]{%
    \vec f \vdot \ten G(\vec x, \vec y) -
    \ten I \diracR3(\vec x - \vec y) \vdot \vec u(\vec x)
  } \dd{V(\vec x)} +
  \int_{I^*} \delim[\big][]{%
    \vec f_\surf \vdot \ten G(\vec x, \vec y) - \ten I_\surf \diracR2(\vec x - \vec y) \vdot \vec u
  } \dd{A(\vec x)} \\
  = \oint_R \bigl\{%
    \uvec n(\vec x) \vdot
      \ten\sigma(\vec x) \vdot \ten G(\vec x, \vec y)
    - \vec u(\vec x) \vdot \bigl[%
      \uvec n(\vec x) \vdot \ten T(\ten G; \vec x, \vec y)
    \bigr]
  \bigr\} \dd{S(\vec x)}.
\end{multline}
We assert that the external force densities $\vec f$ and $\vec f_\surf$ vanish in \cref{eq:BIE-simp-step-0}.
Using the identity \(\int_\Omega \diracR{n}(\vec x - \vec y) \, f(\vec x) \dd^n{\vec x} = \Ind_{\Omega}(\vec y) \, f(\vec y)\), where $f$ and $\Omega$ are arbitrary, \cref{eq:BIE-simp-step-0} becomes
\begin{equation}
  \label{eq:BIE-simp-step-1}
  \begin{aligned}
    &\Ind_{V^*}(\vec y) \, \vec u(\vec y) +
    \Ind_{I^*}(\vec y) \, \ten I_\surf \vdot \vec u(\vec y) \\
    &\quad = \oint_R \bigl\{%
      \uvec n(\vec x) \vdot
      \ten\sigma(\vec x) \vdot \ten G(\vec x, \vec y)
      - \vec u(\vec x) \vdot \bigl[%
        \uvec n(\vec x) \vdot \ten T(\ten G; \vec x, \vec y)
      \bigr]
    \bigr\} \dd{S(\vec x)}.
  \end{aligned}
\end{equation}
The first term on the left-hand side of \cref{eq:BIE-simp-step-1} gives the velocity field (as a function of $\vec y$) whenever $\vec y \in V^*$ (i.e., $\vec y$ is in either $V^*_1$ or $V^*_2$, not including points on the interface) and elsewhere vanishes.
The following term is complementary and vanishes unless $\vec y$ lies right on $I^*$; the surface projection $\ten I_\surf$ has no effect here due to the no-penetration condition \cref{eq:BC-on-I--no-penetration}.

In the limit that $V^* \to V$ and $I^* \to I$, such that the shaded regions in \cref{fig:space} grow to fill the entire domain, with $R^{\text o}_1$ and $R^{\text o}_2$ becoming arbitrarily far away from the colloid, we find that \cref{eq:BIE-simp-step-1} gives the boundary integral representation for the velocity field,
\begin{equation}
  \label{eq:BIE-clean}
  \vec u(\vec x) = -\oint_{S_\text{c}} \ten G(\vec x, \vec y)
  \vdot [\ten\sigma \vdot \uvec n ](\vec y) \dd S(\vec y)
  + \oint_{S_\text{c}} {[\vec u \uvec n]}(\vec y) \odot \ten T (\ten G; \vec y, \vec x)
  \dd S(\vec y),
\end{equation}
where \(S_\text{c} = S_1 \cup S_2\) represents the surface of the colloid and the operator ‘\(\odot\)’ denotes complete contraction of its operands, e.g., \((\ten A \odot \ten B)_{j_1 \dots j_m} = \tsy A_{i_1 \dots i_n} \tsy B_{i_n \dots i_1 j_1 \dots j_m}\) if \(\ten A\) is the tensor of lower rank and \((\ten A \odot \ten B)_{j_1 \cdots j_m} = \tsy A_{j_1 \cdots j_m i_1 \cdots i_n} \tsy B_{i_n \cdots i_1}\) if \(\ten B\) is the tensor of lower rank.
We have exchanged $\vec y$ and $\vec x$ going from \cref{eq:BIE-simp-step-1} to \cref{eq:BIE-clean} to make $\vec x$ be the observation point of $\vec u(\vec x)$ and $\vec y$ be the integration variable.
We have also used the self-adjoint property of $\ten G$ \cref{eq:self-adjoint-G} in the first term on the right-hand side of \cref{eq:BIE-clean}.
The convergence of \cref{eq:BIE-simp-step-1} to \cref{eq:BIE-clean} follows from the decay behavior of $\ten G(\vec x, \vec y)$ and from the quiescent state of the fluid far from the colloid.
\Cref{eq:BIE-clean} is valid as long as the colloid does not deform in a manner that would distort the flat shape of the pinned contact line.

\Cref{eq:BIE-clean} is similar in form and interpretation to the boundary integral equation for Stokes flows that appears in standard textbooks \citep[see, e.g.,][]{Kim1991,Pozrikidis1992}.
Indeed, \cref{eq:BIE-clean} is derived in an analogous manner using the generalized reciprocal relation \cref{eq:R-thm-clean-iface}.
The key property of \cref{eq:BIE-clean} is that, by construction, integrals over the interface itself do not appear because $\ten G$ and $\ten T$ implicitly account for transmission of hydrodynamic stresses through the interface.
This property allows for straightforward generation of the multipole expansion in the following section.

\subsection{Multipole expansion}
\label{ssec:clean-MPX}

To generate a multipole expansion for $\vec u(\vec x)$, we replace $\ten G(\vec x, \vec y)$ and $\ten T(\ten G; \vec x, \vec y)$ in \cref{eq:BIE-clean} with their Taylor series in $\vec y$ about an point on the interface as near as possible to the center of the colloid, which we designate as the origin \(\vec 0\).
This process is slightly complicated by the piecewise nature of \(\ten G\) as \(\vec y\) passes from one side of the interface to the other.
In particular, certain components of \(\grad_{\vec y} \ten G(\vec x, \vec y)\) contain a jump discontinuity over the interface at \(z=0\).
This difficulty is overcome by separating each integral in \cref{eq:BIE-clean} into one over \(S_1\) and another over \(S_2\), so that the integrand is continuous over each of these surfaces.
Letting \(\vec u^{(1)}\) and \(\vec u^{(2)}\) denote the contributions from integration over \(S_1\) and \(S_2\), respectively, we may write the expansion as \(\vec u = \vec u^{(1)} + \vec u^{(2)}\), where
{%
  \newcommand*\sumn{\sum_{n=0}^\infty}
  \newcommand*\intSc[2]{\left( \int_{S_#1} #2 \dd S(\vec y) \right)}
  \newcommand*\ylim[1]{\lim_{\vec y \to \vec 0^#1}}
  \newcommand*\ys{\vec y^{\oprod n}}
  \newcommand*\gradyn{\grad_{\vec y}^{\oprod n}}
  \begin{equation}
    \label{eq:mpx-clean-upper}
    \begin{aligned}
      \vec u^{(1)}(\vec x) =
    -\sumn
      \frac1{n!} \intSc1{[\uvec n \vdot \ten\sigma](\vec y)\,\ys}
      & \odot \left( \ylim+ \gradyn \ten G^\T(\vec x, \vec y) \right)
    \\ +
    \sumn
      \frac1{n!} \intSc1{[\vec u \uvec n](\vec y)\,\ys}
      & \odot \left( \ylim+ \gradyn \ten T(\ten G; \vec y, \vec x) \right)
    \end{aligned}
  \end{equation}
  and
  \begin{equation}
    \label{eq:mpx-clean-lower}
    \begin{aligned}
      \vec u^{(2)}(\vec x) =
      -\sumn
        \frac1{n!} \intSc2{[\uvec n \vdot \ten\sigma](\vec y)\,\ys}
        & \odot \left( \ylim- \gradyn \ten G^\T(\vec x, \vec y) \right)
      \\ +
      \sumn
        \frac1{n!} \intSc2{[\vec u \uvec n](\vec y)\,\ys}
        & \odot \left( \ylim- \gradyn \ten T(\ten G; \vec y, \vec x) \right)
    \end{aligned}
  \end{equation}
Here, \(\ys = \vec y \vec y \cdots \) ($n$ times) denotes the $n$-fold tensor product and \(\gradyn\) similarly denotes the $n$-fold gradient operator.}
Writing \(\ten T\) in terms of \(\ten G\) as
\[
  \tsy T_{ijk}(\ten G; \vec y, \vec x) = -\kd_{ij} P_k(\ten G; \vec y, \vec x)
  + \mu(h) \left(
    \frac{\pd \tsy G_{kj}(\vec x, \vec y)}{\pd y_i} +
    \frac{\pd \tsy G_{ki}(\vec x, \vec y)}{\pd y_j}
  \right)
\]
and collecting terms in \(\ten G\), \(\gradWRT{y} \ten G\), and so on for higher-order gradients of \(\ten G\), we arrive at the multipole expansion,
\begin{subequations}
  \label{eq:mpx-clean}
  \begin{equation}
    \vec u(\vec x) = \vec u^\text{m0}(\vec x) + \vec u^\text{m1}(\vec x) + \vec u^\text{m2}(\vec x) + \text{h.o.t},
    \tag{\ref*{eq:mpx-clean}}
  \end{equation}
  where \(\vec u^\text{m0}\) is the force-monopole (zeroth) moment, \(\vec u^\text{m1}\) is the force-dipole (first) moment, \(\vec u^\text{m2}\) is the quadrupole (second) moment, and so on for higher-order terms (h.o.t.).
  In particular, these first three moments are given by
  \newcommand\args[1]{\vec x, \vec 0^{#1}}
  \begin{align}
    u^\text{m0}_i(\vec x) &= F^{(1)}_i \tsy G_{ij}(\args+) + F^{(2)}_i \tsy G_{ij}(\args-)
    \label{eq:mpx-clean-m0} \\
    u^\text{m1}_i(\vec x) &=
      \tsy D^{(1)}_{jk} \frac{\pd \tsy G_{ij}}{\pd y_k}(\args+) +
      \tsy D^{(2)}_{jk} \frac{\pd \tsy G_{ij}}{\pd y_k}(\args-)
    \label{eq:mpx-clean-m1} \\
    u^\text{m2}_i(\vec x) &=
      \tsy Q^{(1)}_{jkl} \frac{\pd^2 \tsy G_{ij}}{\pd y_l \pd y_k}(\args+) +
      \tsy Q^{(2)}_{jkl} \frac{\pd^2 \tsy G_{ij}}{\pd y_l \pd y_k}(\args-),
    \label{eq:mpx-clean-m2}
  \end{align}
\end{subequations}
where \(\vec F^{(\nu)}\), $\ten D^{(\nu)}$, and $\ten Q^{(\nu)}$ are the monopole, dipole, and quadrupole coefficients for fluid \(\nu \in \{1,2\}\), respectively.
The shorthand notation \(\vec 0^+\) indicates the limit as \(\vec y\) approaches \(\vec 0\) from above the interface (i.e., from fluid 1).
Similarly, \(\vec 0^-\) indicates the limit as \(\vec y\) approaches \(\vec 0\) from below.
In \cref{eq:mpx-clean-m1}, we have assumed, for simplicity, that the colloid does not grow or shrink in volume so that there is no source or sink flow from the origin.
Note that if the colloid is wholly immersed in one fluid, then the multipole coefficients for the other fluid vanish.

At distances far enough from the colloid that points on the colloid surface are virtually indistinguishable from $\vec 0$, \(|\vec x| \gg a \), the leading terms of \cref{eq:mpx-clean} closely approximate $\vec u(\vec x)$.
Recall that \(\ten G(\vec x, \vec y) \sim {|\vec x|}^{-1}\) for \(|\vec x| \gg |\vec y|\).
It follows that \(\vec u^\text{m0}(\vec x) \sim r^{-1}\), where \(r = |\vec x|\).
Each successive multipole moment involves a higher-order gradient of $\ten G$.
Thus, \(\vec u^\text{m1}(\vec x) \sim r^{-2}\), \(\vec u^\text{m2}(\vec x) \sim r^{-3}\) and so on for higher-order moments.
The lowest-order term with a non-zero coefficient dominates the far-field flow.
This behavior is analogous to that of the multipole expansion for objects in a bulk fluid.

\subsubsection{Monopole moment}

The monopole moment corresponds to a point force exerted at the interface, which follows intuitively from the fact that at large distances \(r \gg a\), the colloid is indistinguishable from a single point at the interface.
The functional form of the flow is therefore just that of the Green's function \(\ten G\).
The prefactors appearing in \cref{eq:mpx-clean-m0} are given by
\begin{equation}
  \label{eq:monopole-coeff-clean}
  \vec F^{(\nu)} = -\int_{S_\nu} \ten\sigma \vdot \uvec n \dd S,
\end{equation}
which is the force exerted on fluid \(\nu \in \{1,2\}\) due to motion of the colloid.
There is no need to keep the separate limits on the right-hand side of \cref{eq:mpx-clean-m0} because \(\ten G(\vec x, \vec y)\) is continuous as \(\vec y\) is moved across the interface for fixed \(\vec x\).
This property is not immediately obvious given the potential viscosity difference between the fluids.
Recall, however, the boundary condition \cref{eq:iface-kinematics-G} that demands continuity of \(\ten G\) as \(\vec x\) is brought across the interface for fixed \(\vec y\).
Since \(\ten G\) is self-adjoint \cref{eq:self-adjoint-G}, continuity in \(\vec x\) implies continuity in \(\vec y\).
Indeed, one may verify directly that all three cases in \cref{eq:Gfun-clean} are redundant for \(h \to 0^\pm\).

\Cref{eq:Gfun-clean} in \cref{eq:mpx-clean-m0} yields the monopole moment as
\begin{equation}
  \label{eq:monopole-moment}
  u^\text{m0}_i(\vec x)
  = \frac{1}{\bar\mu} F_k \delta^\para_{kj} \tsy J_{ij}(\vec x),
\end{equation}
where \(\vec F = \vec F^{(1)} + \vec F^{(2)}\) is the total force exerted on both fluids.
\Cref{eq:monopole-moment} shows that \(\vec u^\text{m0}\) is indistinguishable from a Stokeslet in an unbounded fluid of viscosity \(\bar\mu\) associated with the effective force \(\vec F \vdot \ten I_\surf\).
The component of \(\vec F\) normal to the interface does not contribute to the flow at leading order due to the presence of the interface.
The “viscosity-averaged” Stokeslet represented by \cref{eq:monopole-moment} possesses an axis of symmetry lying in the interfacial plane.
The tangential shear stress therefore vanishes at \(z = 0\), and \cref{eq:iface-stress-bal-clean} is trivially satisfied.
More generally, we will find that any mode with mirror symmetry of the velocity field about the interfacial plane has this property and is therefore a viscosity-averaged flow.

\subsubsection{Dipole moment}

The dipole moment is the flow generated by a pair of opposite point forces that are displaced by an infinitesimal distance, or, more generally, a linear combination of such force doublets.
The functional form of this mode is given by \(\gradWRT{y} \ten G(\vec x, \vec y)\) in the limit that \(\vec y\) approaches $\vec 0$ from either side of the interface.
Its prefactor for phase $\nu$ is given by
\begin{equation}
  \label{eq:dipole-coeff-clean}
  \ten D^{(\nu)} = \int_{S_\nu} \left[ -(\ten\sigma \vdot \uvec n) \vec y
  + \mu_\nu (\vec u \uvec n + \uvec n \vec u) \right] \dd{S(\vec y)},
\end{equation}
which we decompose as
\begin{equation}
  \label{eq:dipole-coeff-clean-decomp}
  \tsy D^{(\nu)}_{jk} = \tsy S^{(\nu)}_{jk} + \frac12 \permut_{jkl} L^{(\nu)}_l + \frac13 \tsy D_{ii}^{(\nu)} \kd_{jk}
\end{equation}
where $\ten\permut$ is the permutation tensor.
Here, the irreducible tensor \(\tsy S^{(\nu)}_{jk} = \frac12 ( \tsy D^{(\nu)}_{jk} + \tsy D^{(\nu)}_{kj} ) - \frac13 \tsy D^{(\nu)}_{ii} \kd_{jk}\) is associated with extensional (or contractile) stresses on the fluid, i.e., the stresslet at the interface, and \(\vec L^{(\nu)}\) gives the torque exerted by the colloid on fluid $\nu$.
The total torque exerted by the colloid on the outside system is therefore
\begin{equation*}
  \vec L = \vec L^{(1)} + \vec L^{(2)}
  = \ten\permut \odot \lr(){\ten D^{(1)} + \ten D^{(2)}}
  = -\lr(){\int_{S_1} + \int_{S_2}} \vec y \times (\ten\sigma \vdot \uvec n) \dd{S(\vec y)}.
\end{equation*}
The last term of \cref{eq:dipole-coeff-clean-decomp} is associated with an isotropic stress, which cannot produce flow due to fluid incompressibility \cref{eq:incompressible-G}.
Thus, it makes no contribution to \(\vec u^\text{m1}\).

We may rewrite \cref{eq:mpx-clean-m1} as
\begin{multline}
  \label{eq:um1-clean-expanded}
  u^{\text{m1}}_i(\vec x) =
  \lr(){\tsy D^{(1)}_{\ga\gb} + \tsy D^{(2)}_{\ga\gb}} \frac{\pd \tsy G_{i\alpha}}{\pd y_\beta}(\vec x, \vec 0)
  + \tsy D^{(1)}_{\alpha 3} \frac{\pd \tsy G_{i\alpha}}{\pd h}(\vec x, \vec 0^+)
  + \tsy D^{(2)}_{\alpha 3} \frac{\pd \tsy G_{i\alpha}}{\pd h}(\vec x, \vec 0^-) \\
  + \lr(){\tsy D^{(1)}_{3\beta} + \tsy D^{(2)}_{3\beta}}
  \frac{\pd \tsy G_{i3}}{\pd y_\beta}(\vec x, \vec 0)
  + \lr(){\tsy D^{(1)}_{33} + \tsy D^{(2)}_{33}}
  \frac{\pd \tsy G_{i3}}{\pd h}(\vec x, \vec 0),
\end{multline}
where we introduce the convention that Greek tensor indices, here \(\alpha \in \{1, 2\}\) and \(\beta \in \{1, 2\}\), only run over the axes parallel to the interface.
We have combined the left and right limits in the first, penultimate, and last terms of \cref{eq:um1-clean-expanded} because gradients of \(\ten G\) parallel to the interface are continuous across the interfacial plane by \cref{eq:iface-kinematics-G}.
In the case of the last term, continuity follows from \cref{eq:incompressible-G,eq:iface-kinematics-G,eq:self-adjoint-G}, which give
\begin{equation}
  \label{eq:G33-continuity-proof}
  \lr[]{\frac{\pd \tsy G_{i3}(\vec x, \vec y)}{\pd h}}_I =
  \lr[]{\frac{\pd \tsy G_{3i}(\vec y, \vec x)}{\pd h}}_I =
  \lr[]{\frac{\pd \tsy G_{\alpha i}(\vec y, \vec x)}{\pd y_\alpha}}_I = 0,
\end{equation}
where the usual roles of \(\vec x\) and \(\vec y\) are reversed in the last two equalities.
Furthermore, the penultimate term of \cref{eq:um1-clean-expanded} vanishes because \(\tsy G_{i3}(\vec x, \vec 0) = 0\) by \cref{eq:iface-kinematics-G,eq:self-adjoint-G}.
We cannot similarly combine limits from the second and third terms of \cref{eq:um1-clean-expanded} because \(\lr[]{\pd \tsy G_{i\alpha} / \pd h}_I \neq 0\);
the tangential stress balance \cref{eq:iface-stress-bal-G} requires that
\begin{equation}
  \label{eq:tangential-stress-bal-G}
  \newcommand*\xlim[1]{\lim_{\vec x \to \vec 0^{#1}}}
  \mu_1 \xlim+ \frac{\pd \tsy G_{\alpha k}(\vec x, \vec y)}{\pd z} -
  \mu_2 \xlim- \frac{\pd \tsy G_{\alpha k}(\vec x, \vec y)}{\pd z} = 0.
\end{equation}
However, \cref{eq:self-adjoint-G,eq:tangential-stress-bal-G} relate these limits as
\begin{equation}
  \label{eq:null-dipole}
  \newcommand*\ylim[1]{\lim_{\vec y \to \vec 0^{#1}}}
  \mu_1 \ylim+ \frac{\pd \tsy G_{k \ga}(\vec x, \vec y)}{\pd h} =
  \mu_2 \ylim- \frac{\pd \tsy G_{k \ga}(\vec x, \vec y)}{\pd h} =
  -\frac{\mu(-z)}{\bar\mu} (\kd^\para_{\ga j} n_k + \kd^\para_{\ga k} n_j)
      \frac{\pd \tsy J_{ij}(\vec x)}{\pd x_k},
\end{equation}
where the last equality follows from differentiation of \cref{eq:Gfun-clean}.

Combining \cref{eq:um1-clean-expanded} with \cref{eq:dipole-coeff-clean-decomp,eq:null-dipole} gives
\begin{equation}
  \label{eq:dipole-moment}
    u^\text{m1}_i (\vec x) = -\frac1{\bar\mu} \left(
      \tsy S^\para_{\ga\gb} +
      \frac12 \permut_{\ga\gb3} L_3 -
      S^\perp \kd^\para_{\ga\gb}
    \right) \frac{\pd \tsy J_{i\ga}(\vec x)}{\pd x_\gb} -
    \frac{A_\ga}{\mu(z)} \left(
    \kd^\para_{\ga j} n_k + \kd^\para_{\ga k} n_j
    \right) \frac{\pd \tsy J_{ij}(\vec x)}{\pd x_k},
\end{equation}
where
\begin{subequations}
  \label{eqs:clean-dipole-coeffs}
  \begin{align}
    \tsy S^\para_{\ga\gb} &=
    \tsy S^{(1)}_{\ga\gb} + \tsy S^{(2)}_{\ga\gb} -
    \frac12 \left( \tsy S^{(1)}_{\gc\gc} + \tsy S^{(2)}_{\gc\gc} \right) \kd^\para_{\ga\gb},
    \label{eq:dipole-coeff-para} \\
    S^\perp &=
    \tsy S^{(1)}_{33} + \tsy S^{(2)}_{33} - \frac12 \left( \tsy S^{(1)}_{\gc\gc} + \tsy S^{(2)}_{\gc\gc} \right)
    \label{eq:dipole-coeff-perp}
    \\ 
    \bar\mu A_\ga &= \mu_2 \tsy D^{(1)}_{\ga 3} + \mu_1 \tsy D^{(2)}_{\ga3} =
    \mu_2 \lr(){\tsy S^{(1)}_{\ga 3} + \frac12 \permut_{\ga 3 \gb} L^{(1)}_\gb} +
    \mu_1 \lr(){\tsy S^{(2)}_{\ga 3} + \frac12 \permut_{\ga 3 \gb} L^{(2)}_\gb}.
    \label{eq:dipole-coeff-asym}
  \end{align}
\end{subequations}
\Cref{eq:dipole-moment,eqs:clean-dipole-coeffs} show that the dipole at a clean interface can be conveniently represented in terms of \(\grad \ten J\).

\begin{figure}
  \centering
  \includegraphics{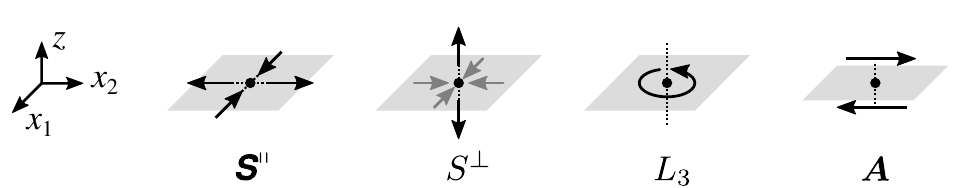}
  \caption{%
    Singularity diagrams corresponding to each of the terms in \cref{eq:dipole-moment}.
    Arrows indicate distributions of point forces or torques and the gray shaded region indicates the interface. 
  }%
  \label{fig:singularities}
\end{figure}

Each term in \cref{eq:dipole-moment} makes a distinct contribution to the interfacial dipole moment.
We call each contribution by its (tensorial) prefactor and graphically represent them as distributions of Stokes singularities in \cref{fig:singularities}.
The \(\ten S^\para\) mode, given by \cref{eq:dipole-coeff-para}, is a viscosity-averaged stresslet associated with extensional stresses produced by the colloid in the interfacial plane.
Similarly, the \(S^\perp\) mode, given by \cref{eq:dipole-coeff-perp}, is the viscosity-averaged stresslet perpendicular to the interface.
Furthermore, \(\tsy S^{(\nu)}_{33} = -\tsy S^{(\nu)}_{11} - \tsy S^{(\nu)}_{22}\) because \(\ten S^{(\nu)}\) is traceless, so \(S^\perp\) accounts for extensional stress perpendicular to the interface and planar compression of the interface.
The $L_3$ mode in \cref{eq:dipole-moment} is a viscosity-averaged rotlet, or point torque, about the \(z\) axis of strength \(L_3\).
These viscosity-averaged flows exhibit mirror symmetry of the velocity field about $z=0$.
Therefore, the tangential shear stress due to these modes vanishes on the interface, as is the case for the monopole moment.

The \(\vec A\) mode corresponds to a force dipole where the forces act parallel to the interface and are displaced from one another along the axis normal to the interface.
This mode does not produce a viscosity-averaged flow.
Instead, the flow speed in one phase differs from that in the opposite phase by a factor of the viscosity ratio; intuitively, the flow is slower in the more viscous phase.
On the interface, the fluid velocity vanishes.
We see from the last term of \cref{eq:dipole-moment} that the flows in the upper and lower half spaces are equivalent to effective stresslets in an unbounded fluid for $z>0$ and $z<0$, respectively.

Quadrupolar and higher-order terms of \cref{eq:mpx-clean} can be similarly decomposed into two subsets of modes; one whose tangential stress vanishes at the interface and another whose velocity vanishes at the interface.
Members of the former subset are mirror-symmetric, viscosity-averaged flows and the latter have velocities that differ by in magnitude by the viscosity ratio on either side of the interface.
We do not detail the higher-order modes further. 
The force monopole \cref{eq:monopole-moment} and force dipole \cref{eq:dipole-moment} describe the leading-order flows due to driven and active colloids, respectively.
In many cases, we can infer these modes based on the geometry of a given colloidal particle and its configuration with respect to the interface.

\subsection{Discussion}

\subsubsection{Driven colloids}

For colloids driven by an external force \(\vec F_\text{ext}\) with a non-zero component parallel to the interface, the monopole moment---a viscosity averaged Stokeslet---is the leading-order far-field flow.
The strength of this effective Stokeslet is simply \(\ten I_\surf \vdot \vec F_\text{ext}\), regardless of whether the colloid is adhered or adjacent to the interface.
An interesting special case occurs when $\vec F_\text{ext}$ acts purely perpendicular to the interface.
For an adhered colloid, this force generates no motion of the colloid---or the fluid---due to the pinned contact line.
However, motion will result if the colloid is instead adjacent to the interface.
In this case, $\vec u^{\text m0}$ still vanishes by \cref{eq:monopole-moment}, so the dipole becomes the leading-order mode.
For instance, consider a colloid fully immersed in fluid 1 whose center is located a small distance $\delta$ from the interface.
This colloid is acted upon by the force \(\vec F_\text{ext} = F_3 \bvec_3\), which drives it in rigid-body motion.
Recall that we have expanded $\vec u$ into multipoles with respect to the origin point $\vec 0$ on the interface, and each multipole prefactor is therefore ‘measured’ with respect to this point.
Letting \(\vec y = \vec y' + \delta \bvec_3\) in \cref{eq:dipole-coeff-clean}, where \(\vec y'\) is the displacement vector from the center of the colloid, we find
\begin{equation}
  \label{eq:dipole-w-offset}
  \ten D^{(1)} = -\int_{S_1} (\ten\sigma \vdot \uvec n) (\vec y' + \delta \bvec_3) \dd{S}(\vec y')
  = \ten D_\text{c} + \delta F_3 \bvec_3 \bvec_3,
\end{equation}
where $\ten \tsy D_{\text c}$ is the dipole strength as measured from the colloid center.
Thus, the external force on the colloid contributes a factor of $\delta F_3 \bvec_3 \bvec_3$ to $\ten D^{(1)}$ (or a factor of $\delta F_3$ to \(S^\perp\)).
If the characteristic size of the colloid $a$ is small compared with $\delta$, then we expect \(\delta F_3 \gg |\ten \tsy D_{\text c}|\).
Otherwise, when $\delta \sim a$, contributions from $\ten D_\text{c}$ are generally significant and are sensitive to particle geometry, its distance to the interface, and the viscosity ratio.

An external torque $\vec L_\text{ext}$ on the colloid also drives flow.
First, consider a torque about the $z$-axis, \(\vec L_\text{ext} = L_{\text{ext},3} \bvec_3\).
This torque is balanced hydrodynamically whether or not the colloid is adhered to the interface because the contact line does not resist rotation about the \(z\) axis.
Thus, \(L_3 = L_{\text{ext},3}\).
The $L_3$ mode of \cref{eq:dipole-moment} induces a viscosity-averaged rotlet.
For colloids that are axisymmetric about the $z$-axis, this is the only non-vanishing mode of \cref{eq:dipole-moment};
it is readily shown that, in this case, \(\tsy S^\para_{jk} = S^\perp = A_\ga = 0\).
For general colloid geometries, these coefficients are generally non-zero, so an external torque potentially produces all of the modes represented by \cref{eq:dipole-moment}.
We may also consider an external torque parallel to the interface.
If the colloid is adhered to the interface, this torque does not produce flow due to the pinned contact line.
For an adjacent colloid immersed in either fluid, the colloid is able to rotate, and we see from \cref{eq:dipole-coeff-asym} that the $\vec A$ mode is produced.
This mode may be accompanied by other dipolar modes that are linearly coupled to the resulting motion of the colloid.

\subsubsection{Active colloids}

\begin{figure}
  \centering
  \includegraphics[width=\linewidth]{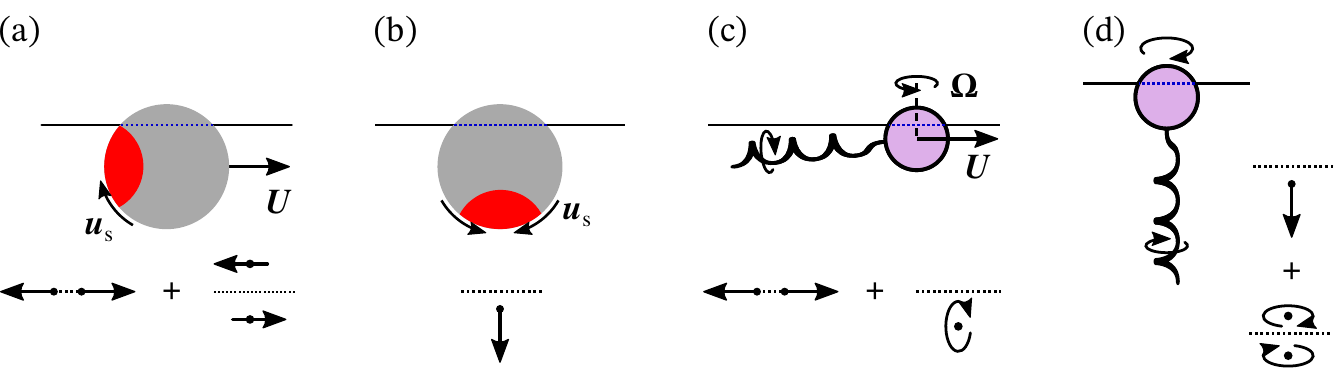}
  \caption{%
    We depict some examples of active colloids at interfaces.
    Panels (a,b) illustrate an active colloidal particle adhered to an interface, which possesses an active cap that generates a phoretic slip velocity $\vec u_\surf$ along its surface.
    In (a), the horizontal particle orientation leads to in-plane swimming at velocity $\vec U$.
    In (b), the same active particle is adhered vertically to the interface as to instead ‘pump’ fluid.
    Here, the particle cannot reorient to swim forward due to a pinned contact line.
    Panels (c,d) illustrate a bacterium also in swimming and pumping configurations.
    Thrust is generated by a rotating flagellum, which also produces a torque.
    In (c), this torque is balanced by contact-line pinning, so there is a net hydrodynamic torque exerted on the fluid below the interface.
    For the vertically adhered bacterium (d), the hydrodynamic torque on the upper and lower fluid must vanish, since the body of the bacterium is free to counterrotate about the \(z\) axis.
    The singularity diagrams next to each illustration give minimal “point-force” models describing to the leading-order flows these active colloids are expected to generate.
    The arrows represent the orientation of these forces or torques (circular arrows) relative to the interface (dashed line).
  }%
  \label{fig:active-colloid-configs}
\end{figure}

\begin{figure}
  \centering
  \includegraphics[width=\linewidth]{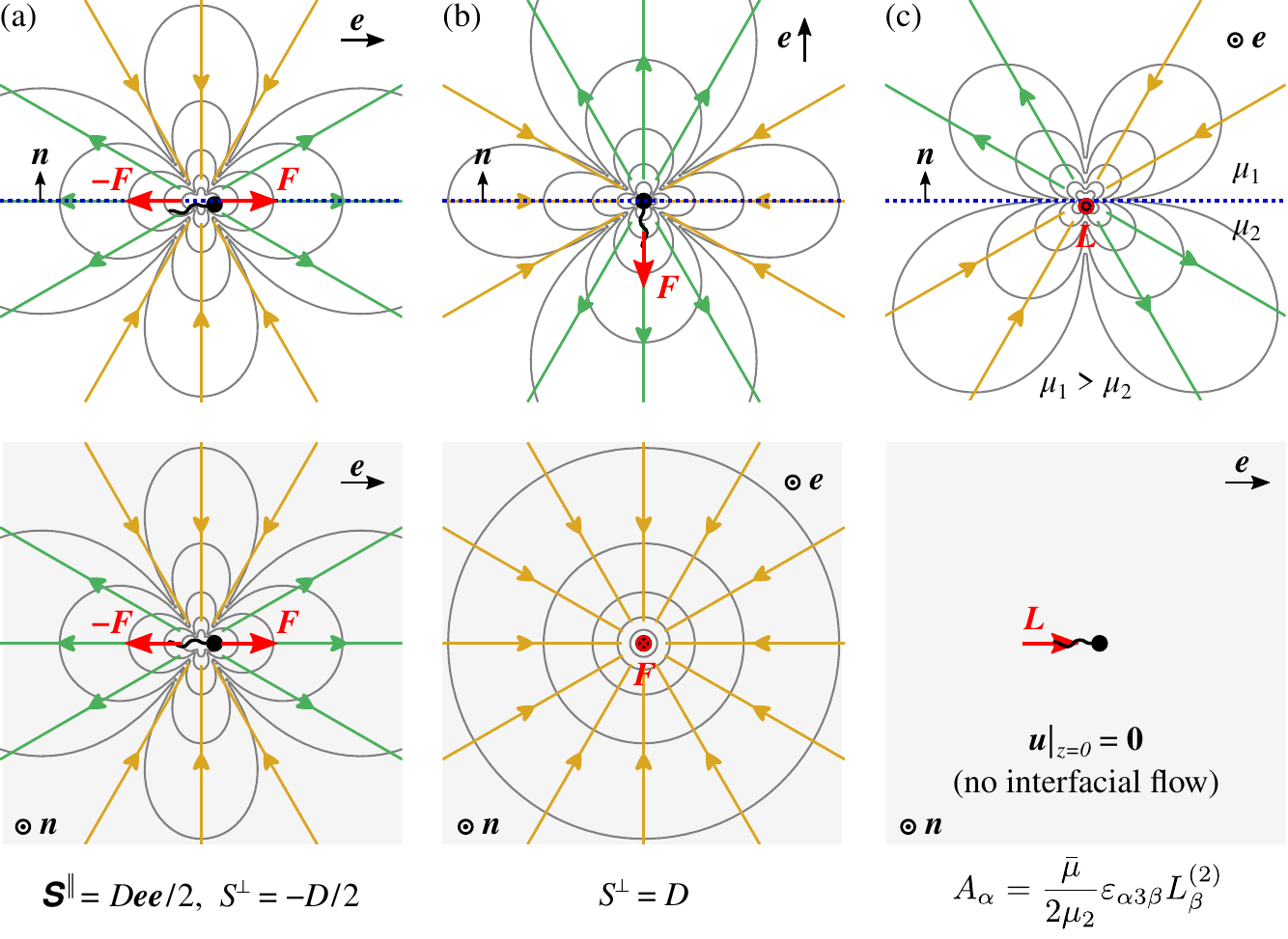}
  \caption{%
    Dipolar hydrodynamic modes for an active colloid at a clean interface.
    The upper panels are side views that depict the flow in a cross-section intersecting the colloid, in which the blue dotted line indicates the location of the interface.
    The lower panels are top-down views of the same flows on the interface.
    Forms of the prefactors from \cref{eq:dipole-moment} associated with each mode are indicated at the bottom.
    The streamlines indicate the flow disturbance due to the colloid, and the gray lines are contours of constant \(|\vec u|\).
    The vector \(\vec e\) represents the alignment of the swimmer.
    (a) Force dipole (stresslet) mode expected for a swimmer moving parallel to the interface.
    The configuration of the swimmer is like that in \cref{fig:active-colloid-configs}(a) or (c).
    (b) Stresslet due to an active colloid pinned at the interface, with a configuration as illustrated in \cref{fig:active-colloid-configs}(b) or (d).
    Modes (a,b) are the same as the force dipole in a bulk fluid with viscosity $\bar\mu$ and are axisymmetric about the swimmer alignment axis.
    (c) Flow due to a point torque $\vec L$ on the lower fluid, just below the interface, where $\vec L$ is parallel to the interface.
    Such a flow is expected for certain active colloids such as the bacterium illustrated in \cref{fig:active-colloid-configs}(c).
    This mode is associated with asymmetry in the activity and/or geometry of the colloid about the interfacial plane, as detailed in \cref{sssec:symmetry}.
  }%
  \label{fig:clean-stresslets}
\end{figure}

Active colloids self-propel absent external forces or torques.
For many kinds of active colloids, self-propulsion is generated by some active, thrust-producing part of the colloid that drives the remaining passive part, as illustrated in \cref{fig:active-colloid-configs}; spatial separation of thrust and drag on the object generate a hydrodynamic dipole.
Therefore, in a bulk fluid, an appropriate far-field model of an active colloid is that of a stresslet along the axis of swimming \citep{Lauga2009}, which produces the velocity field
\begin{equation}
  \label{eq:bulk-axi-dipole}
  \vec u^\text{S}(\vec e; \vec x) = -\frac{D}{\mu_\text{b}} \vec e (\vec e \vdot \grad) \ten J(\vec x),
\end{equation}
where $D$ is the strength of the force dipole, $\mu_\text{b}$ is the viscosity of the bulk fluid, and \(\vec e\) is a unit vector indicating the swimmer alignment.
A similar model is sensible for an active colloid swimming parallel to the interface as illustrated in \cref{fig:active-colloid-configs}(a,c).
Indeed, the same velocity field as \cref{eq:bulk-axi-dipole} is produced by setting \(\ten S^\para = D \vec e \vec e / 2\) and \(S^\perp = -D/2\) in \cref{eq:dipole-moment}, with \(\bar\mu\) replacing \(\mu_\text{b}\).
The resulting flow profile is illustrated in \cref{fig:clean-stresslets}a.

By instead setting \(S^\perp = D\) and \(\ten S^\para = \vec 0\) in \cref{eq:dipole-moment}, one obtains the same flow profile albeit rotated by \SI{90}{\degree}.
This pure-\(S^\perp\) mode is expected of active colloids trapped perpendicular to the interface, \(\vec e = \vec n\), as depicted in \cref{fig:active-colloid-configs}(b,d).
The colloid cannot self-propel in this configuration due to the pinned contact line, so the apparent stresslet \cref{eq:bulk-axi-dipole} is not due to balancing hydrodynamic thrust and drag.
Instead of swimming, the colloid becomes a fluid pump, resulting in a non-zero net hydrodynamic force on the colloid that is balanced by capillary forces.
A minimal model for this pumping configuration is that of a point force exerted along the \(z\) axis a small distance \(\delta\) from the interface.
While the monopole moment vanishes for a force in this direction, the dipole moment does not due to the small but finite separation of the force from the interface.
The vertical point force gives \(S^\perp = F \delta\) in \cref{eq:dipole-moment}, which is associated with the flow plotted in \cref{fig:clean-stresslets}b.
Viewed in the interfacial plane, this flow is sink like for a pusher (\(S^\perp > 0\)) and source like for a puller (\(S^\perp < 0\)).
A pusher causes surface expansion (\(\divS \vec u > 0\)), as new interface must be created to replace the ‘sink.’
Conversely, a puller causes surface compression.

Another unique feature of active colloids adhered to interfaces is that they may exert a net hydrodynamic torque on the fluid about an axis parallel to the interface.
This torque is balanced by capillary forces at the contact line.
\Cref{fig:active-colloid-configs}c illustrates this scenario for a motile bacterium adhered to the interface by its body and propelled by a rotating flagellum.
The effect of this torque on the far-field flow enters through the coefficient \(\vec A\) in \cref{eq:dipole-moment}.
The resulting flow profile is shown in \cref{fig:clean-stresslets}c.
The presence of this mode potentially discriminates the far-field flow of adhered versus unadhered swimmers; the net torque must vanish for active colloids that are adjacent but not adhered to the interface.
In the case of an adjacent bacterium, counterrotation of the body and flagellum instead produce a torque dipole in the far-field, a member of the higher-order quadrupole moment.
A perpendicular configuration of the bacterium, as in \cref{fig:active-colloid-configs}(d) produces a torque dipole as well because the body may freely counterrotate in in the interface.
As discussed further below, this mode is of particular interest in advective mixing near fluid interfaces, regardless of interfacial mechanics.

\subsubsection{Symmetry and asymmetry about the interfacial plane}
\label{sssec:symmetry}

To conclude this discussion, we return to the motif of two major categories of modes: those which are weighted by the average viscosity, with vanishing tangential stress at the interface, and those whose flow speed on either side of the interface differs by a factor of the viscosity ratio, with vanishing velocity on the interface.
To dipolar order, only the $\vec A$ mode in \cref{eq:dipole-moment} falls into the latter category.
The previous discussion associated \(\vec A\) with a net hydrodynamic torque on the fluid adjacent to the interface about an axis parallel to the interface.
Such torques might arise from active stresses or, for colloids adjacent to the interface, a driving external torque.
However, this mode is not uniquely associated with these torques; from \cref{eq:dipole-coeff-asym}, we see that it also involves the components of the stresslet \(\tsy S^{(\nu)}_{\alpha 3} \).

To gain a better understanding of the $\vec A$ mode, consider a spherical colloid of radius $a$ that is adhered to the interface with a \SI{90}{\degree} contact angle, such that half of the sphere is in each fluid.
We may exactly obtain the flow due to rigid-body motion of this sphere by referencing an auxiliary problem where the sphere instead moves through a fluid of uniform viscosity $\bar\mu$ \citep{Ranger1978,Pozrikidis2007}.
If the sphere translates at velocity $\vec U$ in the $z=0$ plane and rotates with angular velocity $\bvec_3 \Omega_3$, the fluid velocity in the laboratory frame with its origin at the center of the sphere is
\begin{equation}
  \label{eq:sphere-flow}
  \vec u(\vec x) = \vec F \left(1 + \frac{a^2}{6} \nabla^2 \right) \vdot \ten J(\vec x)
  + \frac12 L_3 \bvec_3 \vdot [\grad \times \ten J(\vec x)],
\end{equation}
where \(\vec F = 6 \upi \bar\mu \vec U a\) is the Stokes drag and \(L_3 = 8 \upi \bar\mu \Omega_3\) is the torque.
This velocity field is mirror symmetric about the \(z=0\) plane, so the tangential stress vanishes on \(z=0\).
It follows that \cref{eq:sphere-flow} trivially satisfies \cref{eq:iface-stress-bal-clean} and is therefore also the solution for two fluids of differing viscosities that average to \(\bar\mu\); the flow is independent of the viscosity contrast.
There is a normal stress jump across the interface in this case, but it is inconsequential at small $\numCa$, where the interface remains nearly flat.

The first term of \cref{eq:sphere-flow} comprises a viscosity-averaged Stokeslet and degenerate quadrupole (or source doublet) at the center of the sphere.
Thus, for the sphere described above, the dipole moment vanishes, excepting the viscosity-averaged rotlet described by the $L_3$ mode of \cref{eq:dipole-moment}.
If there is no external torque on the sphere but it translates along, e.g., the \(x\) axis, then we expect a non-zero hydrodynamic torque about the \(y\) axis unless \(\mu_1 = \mu_2\).
One might naively expect this hydrodynamic torque to produce flow, which clearly contributes to \(\vec A\) in \cref{eq:dipole-coeff-asym}.
However, for the velocity field produced by a sphere, the \(\ten S^{(\nu)}\) contribution to \(\vec A\) exactly cancels that from \(\vec L^{(\nu)}\) due to the sphere's symmetry about the interfacial plane.

More generally, we expect a viscosity-averaged flow to result for any driven or active colloid with mirror symmetry about $z=0$.
If the boundary motion is symmetric about $z=0$, then the resulting fluid flow will reflect this symmetry.
Thus, the \(\vec A\) mode only contributes to the flow when there is some degree of asymmetry.
For rigid, driven colloids, this asymmetry may come from an asymmetric colloid shape or adhered configuration with the interface (for a sphere, a contact angle other than \SI{90}{\degree}).
For active colloids, there will likely be asymmetry in activity or boundary motion, especially if the two fluid phases have differing viscosities or chemical properties.
For example, the phoretic swimmer illustrated in \cref{fig:active-colloid-configs}a is expected to produce a leading-order stresslet parallel to the interface due to hydrodynamic thrust and drag (\cref{fig:clean-stresslets}a).
However, we also expect a contribution from the asymmetric mode illustrated by \cref{fig:clean-stresslets}c.
In experiment, contact-line pinning fixes colloids in random configurations at fluid interfaces, so such asymmetric adhered states are likely the norm.

\section{Incompressible interfaces and the role of surface viscosity}
\label{sec:incompressible-interfaces}

As discussed in \cref{sec:governing-eqs-incompressible}, fluid interfaces are typically incompressible due to the inevitable presence of surface-active impurities. 
Because materials accumulate at interfaces, they often act as two-dimensional fluids with their own rheology.
Here, we address incompressible interfaces with zero and finite shear viscosities.

\subsection{Green's function}
\label{ssec:dirty-Green}

We may define a Green's function \(\ten H\) for an incompressible interface that is analogous to that discussed in \cref{ssec:clean-Green} for a clean interface.
The major difference is that the interfacial stress balance \cref{eq:iface-stress-bal-G} is replaced by
\begin{subequations}
  \label{eq:iface-stress-bal-H}
  \begin{align}
    - \gradS \vec\Pi
    + \mu_\surf \laplaceS \ten H
    + \ten I_\surf \vdot \lr[]{\vec n \vdot \ten T(\ten H;\!)}_I &=
    \begin{cases}
      -\ten I_\surf \diracR2(\vec x - \vec y)  &  h = 0      \\
      \vec 0                               &  h \neq 0
    \end{cases}
    \\
    \label{eq:surf-div-H}
    \divS{\ten H} &= 0,
  \end{align}
\end{subequations}
where \(\ten\Pi\) is the (vectorial) surface pressure associated with \(\ten H\), which enforces the surface incompressibility constraint \cref{eq:surf-div-H}.
Thus, \(\ten H\) satisfies \cref{eq:Stokes-G} subject to \cref{eq:iface-kinematics-G} and \cref{eq:iface-stress-bal-H}, with \(\ten G\) replaced by \(\ten H\) in the former two equations.
The coupling of bulk-viscous and surface-viscous effects induce a natural length scale in the problem, the Boussinesq length, which is given by \(\lenBq = \mu_\surf / \bar\mu\).
Bulk-viscous dissipation dominates at distances \(r \gg \lenBq\), and surface-viscous dissipation dominates when \(r \ll \lenBq\).
It is therefore convenient to define the dimensionless Boussinesq number, \(\numBq = \lenBq/a\), which quantifies the relative importance of surface-viscous to bulk-viscous effects for a colloid of characteristic size $a$.

The functional form of \(\ten H\), derived in full by \citet{Blawzdziewicz1999}, is more complicated than that of \(\ten G\) owing to non-trivial interfacial stress balance \cref{eq:iface-stress-bal-H}.
However, like \(\ten G\), \(\ten H\) is self-adjoint (see \cref{sec:app:self-adjointness}) and therefore satisfies
\begin{equation}
  \label{eq:self-adjoint-H}
  \ten H(\vec x, \vec y) = \ten H^\T(\vec y, \vec x).
\end{equation}
We may use this property to determine the multipole expansion at points on the interface requiring only knowledge of \(\ten H(\vec x, \vec y)\) for \(\vec y \in I\), in addition to the previously determined expansion for a clean interface.
Letting \(\ten H^0(\vec s, z) = \ten H(\vec x, \vec 0)\), we find that for a fixed value of $\lenBq$ (see \cref{sec:app:Gfuns} for derivation)
\begin{align}
  \label{eq:Hfun}
  \tsy H^0_{\ga\gb}(\vec s, z) = 
  \frac{1}{4\upi \bar\mu} R_0(\lenBq; s, |z|) \kd_{\ga\gb} +
  \frac{1}{2\upi \bar\mu} R_2(\lenBq; s, |z|) \Irr{\hat s_\ga \hat s_\gb}
\end{align}
and \(\tsy H^0_{3i} = \tsy H^0_{i3} = 0\), where \(\vec s = \ten I_\surf \vdot \vec x = x_1 \bvec_1 + x_2 \bvec_2\), \(s = |\vec s|\), \(\uvec s = |\vec s| / s\), and \(\Irr{\cdot}\) denotes the irreducible (traceless, symmetric) part of the enclosed tensor.
Here, e.g., \(\Irr{\hat s_\ga \hat s_\gb} = \hat s_\ga \hat s_\gb - \frac12 \kd_{\ga\gb}\), where we regard $\uvec s$ as a two dimensional vector and the \(\{\cdot\}_0\) operation as being on a two-dimensional tensor.
The functions \(R_0\) and \(R_2\), given by \cref{eq:rzFn}, depend only on the magnitudes of $\vec s$ and $z$.
\Cref{eq:self-adjoint-H} equips \(\ten H^0\) with dual interpretations; \(\tsy H^0_{ij}(\vec s, z)\, F_j\) is the velocity field due to the point force \(\vec F\) on the interface at \(\vec y = \vec 0\), and \(\tsy H^0_{ij}(\vec s, h)\, F_j\) is the velocity field on the interface induced by the same point force exerted at \(\vec y = h \bvec_3\) (within either of the fluids).

The velocity field represented by \(\ten H^0(\vec s, z)\) is everywhere parallel to the interface.
As noted by \citet{Stone2015}, this feature is generally expected of Stokes flow driven by arbitrary motion of an incompressible plane.
The $z$ velocities of the fluids vanish at the interface as do their $z$ derivatives due to the incompressibility of the interface and the surrounding fluids.
These quantities also vanish as \(|\vec x| \to \infty\).
As a Stokes flow, \(\vec u\) is biharmonic, and hence \(\nabla^4 u_3 = 0 \).
It follows from the homogeneous behavior of $u_3$ on the interface and at infinity that \(u_3 = 0\) everywhere.
The vanishing behavior of \(\tsy H^0_{3j}\) reflects this property.

When the interface is inviscid, \(\lenBq = 0\), \(R_n\) admits the closed form given by \cref{eq:rzFn|Bq=0}, and \cref{eq:Hfun} reduces to
\begin{equation}
  \label{eq:Hfun0|Bq=0}
  \lr.|{\tsy H^0_{\ga\gb}}_{\lenBq=0} = 
  \frac{\kd_{\ga\gb}}{8\upi\bar\mu r} +
  \frac{(r - |z|)^2}{4\upi \bar\mu r s^2} \Irr{\hat s_\ga \hat s_\gb}.
\end{equation}
Thus, Marangoni stresses alone do not change the \(r^{-1}\) rate of decay of the fluid velocity from the singular point at the origin found for a clean interface.
The flow on the interface is purely radial and is given by
\begin{equation*}
  \lr.|{\tsy H^0_{\ga\gb}(\vec x \in I)}_{\numBq=0} = 
  \frac{\hat s_\ga \hat s_\gb}{8\upi\bar\mu r}.
\end{equation*}
For finite \(\lenBq\), \(\ten H\) approaches the form given by \cref{eq:Hfun0|Bq=0} for \(r \gg \lenBq\).
In the opposite limit where \(\lenBq \to \infty\), we recover the equations governing a two-dimensional Stokes flow from \cref{eq:iface-stress-bal-H}, as viscous stresses from the two fluid phases become negligible.
Therefore, \(\ten H^0\) coincides with a Stokeslet in a two dimensional fluid (up to an arbitrary constant) \citep{Saffman1975},
\begin{equation}
  \label{eq:Hfun0|Bq=oo}
  \lr.|{\tsy H^0_{\ga\gb}}_{\lenBq\to\infty} \sim
    \frac{\hat s_\ga \hat s_\gb - \kd_{\ga\gb} \ln s}{4\pi\mu_\surf},
\end{equation}
which is constant in $z$ (as a three-dimensional flow field) and diverges logarithmically as $s$ is made large.
For finite \(\lenBq\), \(\ten H\) approaches the form given by \cref{eq:Hfun0|Bq=oo} for \(r \ll \lenBq\).
Clearly, \cref{eq:Hfun0|Bq=oo} does not satisfy the homogeneous boundary conditions imposed on $\ten H$ as \(r \to \infty\).
Of course, this is Stokes' paradox; for large but finite $\lenBq$, \cref{eq:Hfun0|Bq=oo} is not valid for \(r \gtrsim \lenBq\), where bulk-viscous effects inevitably become important.
Returning to the general expression \cref{eq:Hfun}, we find that the \(R_n\) functions transition the flow field between the surface-viscosity-dominated, logarithmically divergent behavior at distances \(r \ll \lenBq\) from the colloid to the convergent, \(1/r\) decay at distances \(r \gg \lenBq\), where surface viscosity has a negligible effect.

The surface pressure \(\vec\Pi = \vec\Pi(\vec x, \vec y)\) associated with \(\ten H(\vec x, \vec y)\) is independent of \(\lenBq\), as has been previously found \citep{Blawzdziewicz1999,Fischer2006}, and is given in our notation by (see \cref{sec:app:Gfuns} for derivation)
\begin{equation}
  \label{eq:surface-pressure}
  \Pi_i(\vec s, h \bvec_3) =
  \frac{s_i}{4\upi s^2} \lr(){1 - \frac{|h|}{\sqrt{s^2 + h^2}}} -
  \frac{|h| (s_i - h n_i)}{4\upi {(s^2 + h^2)}^{3/2}}.
\end{equation}
\Cref{eq:surface-pressure} shows that \(|\ten\Pi| \sim 1/s\) for $h$ fixed and \(s \gg h\), and \(|\ten\Pi| \sim 1/h\) for $s$ fixed and \(h \gg s\).
For \(h = 0\), \cref{eq:surface-pressure} reduces to the harmonic pressure field \(\ten\Pi(\vec s, \vec 0) = \vec s / 4\upi s^2\).

\subsection{Multipole expansion}

\subsubsection{Expansion of the boundary integral equation}

Using the reciprocal relation \cref{eq:R-thm-incompressible-iface} for two fluids separated by an incompressible interface and following a procedure similar to that described in \cref{ssec:clean-BIE}, we obtain the boundary integral representation for \(\vec u(\vec x)\) as
\begin{multline}
  \label{eq:BIE-incompr}
  u_k(\vec x) = - \oint_{S_\text{c}} {%
    \tsy H_{kj}(\vec x, \vec y) \, [\hat n_i \sigma_{ij}](\vec y)
  }\dd S(\vec y)
  + \oint_{S_\text{c}} {%
    [\hat n_i u_j](\vec y) \, \tsy T_{ijk}(\ten H; \vec y, \vec x)
  }\dd S(\vec y)
  \\
  - \oint_C {%
    \tsy H_{k\beta}(\vec x, \vec y)\, [\hat m_\alpha \varsigma_{\alpha\beta}](\vec y)
  } \dd C(\vec y)
  + \oint_C {%
    [\hat m_\alpha u_\beta](\vec y) \,
    \Sigma_{\alpha\beta k}(\ten H; \vec y, \vec x)
  } \dd C(\vec y),
\end{multline}
where \(\uvec m = \bvec_3 \times \uvec t\), \(C\) is the curve in the \(z=0\) plane that runs along the three-phase contact line (see \cref{fig:space}), and \(\ten\Sigma(\ten H;\!)\) is the surface stress tensor associated with \(\ten H\), which is given by
\begin{equation*}
  \Sigma_{\ga\gb k}(\ten H; \vec y, \vec x) =
  -\kd^\para_{\ga\gb} \Pi_k(\vec y, \vec x) +
  \mu_s \left(
    \frac{\pd \tsy H_{k\gb}(\vec x, \vec y)}{\pd y_\ga} +
    \frac{\pd \tsy H_{k\ga}(\vec x, \vec y)}{\pd y_\gb}
  \right)
\end{equation*}
for \(\vec y \in I\).
\Cref{eq:BIE-incompr} is valid as long as $\lenBq$ is finite.
We require the $1/r$ spatial decay of $\ten H$ at distances \(r \gg L_B\) from the colloid for the integrals in this equation to converge unconditionally.
Comparing \cref{eq:BIE-incompr} to \cref{eq:BIE-clean}, the last two terms of \cref{eq:BIE-incompr} are new and account for Marangoni forces and surface-viscous stresses at the contact line, respectively.

As before, we may generate a multipole expansion for \(\vec u(\vec x)\) by replacing \(\ten H\), \(\ten T(\ten H;\!)\), and \(\ten\Sigma(\ten H;\!)\) in \cref{eq:BIE-incompr} with their respective Taylor series about \(\vec y = \vec 0\), where $\vec 0$ is placed at an appropriate point on the interface.
We may write the expansion as
\(\vec u = \vec u^{(1)} + \vec u^{(2)} + \vec u^{(\text i)}\), where
\begin{align}
  \label{eq:mpx-incompr-upper}
  \vec u^{(1)} &= \text{same as right-hand side of \cref{eq:mpx-clean-upper}
                        with \(\ten G\) replaced by \(\ten H\),}
  \\
  \label{eq:mpx-incompr-lower}
  \vec u^{(2)} &= \text{same as right-hand side of \cref{eq:mpx-clean-lower}
                        with \(\ten G\) replaced by \(\ten H\),}
\end{align}
and
\begin{equation}
  \newcommand*\sumn{\sum_{n=0}^\infty}
  \newcommand*\intC[1]{\left( \int_C #1 \dd C(\vec y) \right)}
  \newcommand*\ys{y_{\gc_1} \cdots y_{\gc_n}}
  \newcommand*\gradSyn[1]{\left.
      \frac{\pd^n #1}{\pd y_{\gc_1} \cdots \pd y_{\gc_n}}
    \right|_{\vec y = \vec 0}}%
  \label{eq:mpx-incompr-iface}
  \begin{aligned}
    u^{(\text i)}_k (\vec x) =
    -&\sum_{n=0}^\infty {%
      \frac1{n!} \intC{[\hat m_\ga \varsigma_{\ga\gb}](\vec y)\,\ys}
      \gradSyn{\tsy H_{k\gb}(\vec x, \vec y \in I)}
    } \\
    +&\sum_{n=0}^\infty {%
      \frac1{n!} \intC{[u_\gb \hat m_\ga](\vec y)\,\ys}
      \gradSyn{\Sigma_{\ga\gb k}(\ten H; \vec y \in I, \vec x)}
    }.
  \end{aligned}
\end{equation}
Collecting terms from \cref{eq:mpx-incompr-upper,eq:mpx-incompr-lower,eq:mpx-incompr-iface}, we may write \(\vec u\) as a multipole expansion analogous to that given by \cref{eq:mpx-clean-lower},
\begin{subequations}
  \label{eq:mpx-incompr}
  \begin{equation}
    \vec u(\vec x) = \vec u^\text{m0}(\vec x) + \vec u^\text{m1}(\vec x) + \vec u^\text{m2}(\vec x) + \text{h.o.t},
    \tag{\ref*{eq:mpx-incompr}}
  \end{equation}
  where
  \newcommand\args{\vec x, \vec 0}
  \newcommand\argslim[1]{\vec x, \vec 0^{#1}}
  \begin{align}
    u^\text{m0}_i(\vec x) &=
    F^{(1)}_j \tsy H_{ij}(\argslim+) +
    F^{(2)}_j \tsy H_{ij}(\argslim-) +
    F^{(\text i)}_\gb \tsy H_{i\gb}(\args)
    \label{eq:mpx-incompr-m0} \\
    u^\text{m1}_i(\vec x) &=
      \tsy D^{(1)}_{jk} \frac{\pd \tsy H_{ij}}{\pd y_k}(\argslim+) +
      \tsy D^{(2)}_{jk} \frac{\pd \tsy H_{ij}}{\pd y_k}(\argslim-) +
      \tsy D^{(\text i)}_{\gb\gc} \frac{\pd \tsy H_{i\gb}}{\pd y_\gc}(\args).
    \label{eq:mpx-incompr-m1}
  \end{align}
\end{subequations}
\Cref{eq:mpx-incompr-m0,eq:mpx-incompr-m1} are analogous to \cref{eq:mpx-clean-m0,eq:mpx-clean-m1}, respectively, and each of these equations contain an additional term that accounts for stresses exerted by the colloid on the interface.
\Cref{eq:mpx-incompr-m1} assumes that the hole in the interface created by the colloid is of constant surface area and that the volume of the colloid is also constant.

\subsubsection{Monopole Moment}

At an incompressible interface, the monopole moment retains its interpretation as the leading-order flow due to a colloid that is driven by an external force.
Indeed, \(\ten H(\vec x, \vec y)\) is continuous as \(\vec y\) is moved across the interface, so \({\vec u}^\text{m0}(\vec x)\) reduces to
\begin{equation}
  \label{eq:monopole-moment-incompr}
  u^\text{m0}_i(\vec x) = F_j \tsy H^0_{ij}(\vec x) =
  \left(
    F^{(1)}_j + F^{(2)}_j + F^{(\text i)}_\beta \kd^\para_{\beta j}
  \right) \tsy H^0_{ij}(\vec x),
\end{equation}
where $\vec F$ is the total force exerted on the surrounding system: the two fluids and the interface.
Compared with the analogous expression for a clean interface \cref{eq:monopole-moment}, the surface-incompressible monopole has the additional contribution to its prefactor given by
\(
  F^{(\text i)}_{\gb} = -\oint_C \hat m_\alpha \varsigma_{\alpha\beta} \dd C,
\)
which is the force exerted on the interface by the colloid at the contact line due to Marangoni and surface-viscous stresses.
Like the clean-interface monopole, the surface-incompressible monopole given by \cref{eq:monopole-moment-incompr} is independent of the viscosity contrast between the two fluids and is mirror symmetric with respect to the \(z=0\) plane.
However, the surface-incompressible monopole lacks the axisymmetry of its clean-interface analogue.
From \cref{eq:monopole-moment-incompr}, we find that \(\vec u^\text{m0}\) only contributes to velocity components parallel to the interface because \(\tsy H^0_{i3} = 0\).
Hence, this mode is associated with ‘lamellar’ fluid motion, a feature not shared with its clean-interface analogue.

\begin{figure}
  \centering
  \includegraphics[width=\linewidth]{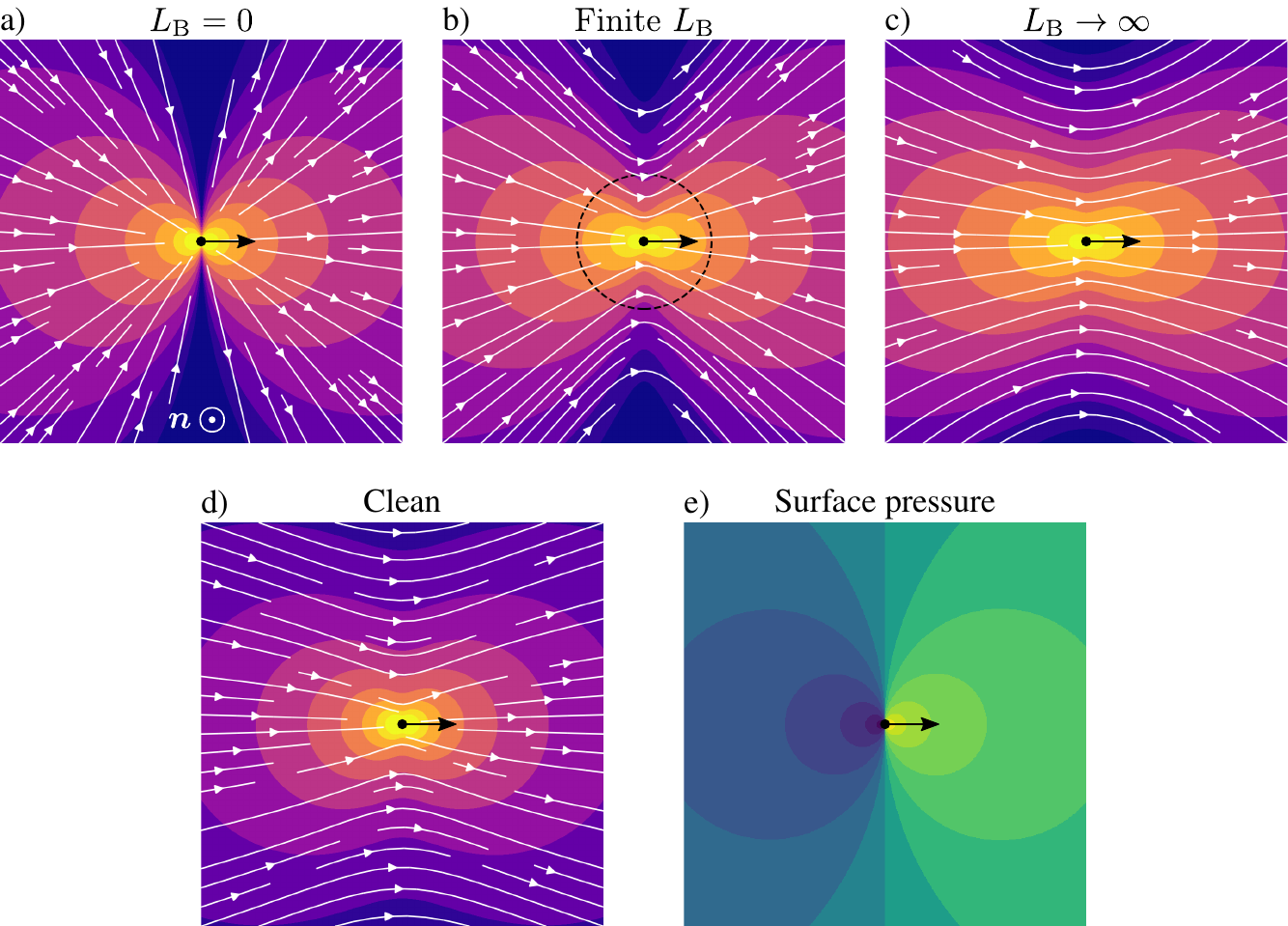}
  \caption{%
    The velocity field on the interfacial plane due to a surface-incompressible force monopole is shown for different values of  $\lenBq$ (a-c).
    The result for a clean interface (d) is also shown for comparison.
    In (b), the distance \(r = \lenBq\) is indicated by the dashed circle.
    The arrows show the location and direction of the point of forcing.
    The surface pressure profile (e) associated with (a-c) is independent of \(\lenBq\).
    The logarithmically spaced contours correspond to velocity magnitude (a-d) and surface pressure (e), where lighter colored contours indicate larger values and darker colors approach zero.
    Panels (a), (b), and (c), represent the leading-order velocity fields due to a colloid driven by an external force for small-, moderate-, and large-\(\numBq\), respectively.
  }%
  \label{fig:monopoles}
\end{figure}

\Cref{fig:monopoles} shows the interfacial velocity and surface pressure profiles of the force monopole \cref{eq:monopole-moment-incompr}.
For \(\lenBq = 0\) (\cref{fig:monopoles}a), the flow is quite different than that for a clean interface (\cref{fig:monopoles}d), although in both cases the velocity magnitude decays as $1/r$.
Here, the flow is modified from that at a clean interface by Marangoni stresses alone.
\Cref{fig:monopoles}a is representative of the leading-order flow for a driven colloid whose characteristic size far exceeds $\lenBq$ (\(\numBq \ll 1\)), where surface-viscous forces are weak everywhere on the interface.
The moderate-\(\numBq\) case is shown in \cref{fig:monopoles}b.
The more complicated flow pattern owes to the interplay between bulk-viscous, surface-viscous, and Marangoni stresses, which are all comparable in magnitude for \(r \sim \lenBq\).
\Cref{fig:monopoles}c shows the large-\(\numBq\) regime, where dissipation is dominated by the surface viscosity except at very large distances of \(r \geq O(a\numBq)\).
For smaller $O(a)$ distances, the flow profile has the same angular dependence about the $z$ axis as is found for a clean interface, but the magnitude of the velocity behaves as \(-\ln(r)\) rather than as \(1/r\).
The surface pressure field associated with \(\vec u^\text{m0}\), \(\vec F \vdot \vec\Pi|_{h=0}\) (\cref{fig:monopoles}e), does not vary with $\lenBq$.

\subsubsection{Dipole Moment}

Similar to the monopole moment, the dipole moment has an additional contribution due to interfacial stresses given by the final term in \cref{eq:mpx-incompr-m1}, whose prefactor is
\begin{equation}
  \tsy D^{(\text i)}_{\beta\gamma} = \oint_C \left[
    -\hat m_\alpha \varsigma_{\alpha\beta} y_\gamma
    + \mu_\surf (\hat m_\beta u_\gamma + u_\gamma \hat m_\beta)
  \right] \dd C(\vec y),
\end{equation}
where $\vec m$, $\ten\varsigma$, and $\vec u$ are regarded as functions of $\vec y$.
We may rewrite \cref{eq:mpx-incompr-m1} as
\begin{equation}
  \label{eq:um1-incompr-expanded}
  u^{\text{m1}}_i(\vec x) =
  \lr(){\tsy D^{(1)}_{\ga\gb} + \tsy D^{(2)}_{\ga\gb} + \tsy D^{(\text i)}_{\ga\gb}}
    \frac{\pd \tsy H_{i\alpha}}{\pd y_\beta}(\vec x, \vec 0)
  + \tsy D^{(1)}_{\alpha 3} \frac{\pd \tsy H_{i\alpha}}{\pd h}(\vec x, \vec 0^+)
  + \tsy D^{(2)}_{\alpha 3} \frac{\pd \tsy H_{i\alpha}}{\pd h}(\vec x, \vec 0^-).
\end{equation}
\Cref{eq:um1-incompr-expanded} omits terms involving derivatives of \(\tsy H_{i3}\), which vanish because the interface is impenetrable and because incompressibility of the interface and the surrounding fluid imply that
\begin{equation}
  \label{eq:H-deriv-props}
  0 = \lr.|{\frac{\pd \tsy H_{\alpha j}(\vec x, \vec y)}{\pd x_\alpha}}_{z = 0}
  = \lr.|{\frac{\pd \tsy H_{3j}(\vec x, \vec y)}{\pd z}}_{z = 0}
  = \lr.|{\frac{\pd \tsy H_{j3}(\vec x, \vec y)}{\pd h}}_{h = 0},
\end{equation}
where the final equality follows from \cref{eq:self-adjoint-H}.

We may decompose \(\ten D^{(1)}\) and \(\ten D^{(2)}\) into irreducible tensors as before (see eq.~\ref{eq:dipole-coeff-clean-decomp}).
A similar decomposition of the two-dimensional tensor \(\ten D^{(\text i)}\) is given by
\begin{equation}
  \label{eq:iface-dipole-coeff-decomp}
  \tsy D^{(\text i)}_{\ga\gb}
  = \tsy S^{(\text i)}_{\ga\gb}
  + \frac12 \permut_{\ga\gb3} L^{(\text i)}
  + \frac12 \kd_{\ga\gb} \tsy D^{(\text i)}_{\gc\gc},
\end{equation}
where the irreducible part of \(\ten D^{(\text i)}\) is given by
\[
  \tsy S^{(\text i)}_{\ga\gb} =
  \frac12 \lr(){\tsy D^{(\text i)}_{\ga\gb} + \tsy D^{(\text i)}_{\gb\ga}}
  - \frac12 \kd_{\ga\gb} \tsy D^{(\text i)}_{\gc\gc},
\]
which represents the stresslet on the interface due to forces on the contact line.
Similarly, the pseudoscalar
\(
  L^{(\text i)} = -\bvec_3 \vdot \oint_C {%
    \vec y \times [\uvec m \vdot \ten\varsigma](\vec y)
  } \dd C(\vec y)
\)
is the torque (about the \(z\) axis) exerted on the interface by the colloid.
The total torque exerted on the surrounding system (both fluids and the interface) is therefore \(\vec L = \vec L^{(1)} + \vec L^{(2)} + L^{(\text i)} \bvec_3\).
From \cref{eq:iface-stress-tensor-incompr}, it is readily shown that the surface pressure makes no contribution to \(L^{(\text i)}\), and therefore \(L^{(\text i)} = 0\) if \(\mu_\surf = 0\).

Finally, \cref{eq:surf-div-H,eq:self-adjoint-H,eq:H-deriv-props} imply that
\( [{\pd \tsy H_{j \alpha}(\vec x, \vec y)} / {\pd y_\alpha}]_{h = 0} = 0 \).
Comparing this result with \cref{eq:um1-incompr-expanded} reveals that the ‘surface’ traces of \(\ten D^{(1)}\), \(\ten D^{(2)}\), and \(\ten D^{(\text i)}\), i.e., \(\tsy D^{(\nu)}_{\gc\gc}\), are of no dynamical significance because the interface is incompressible.
It follows from the bulk incompressibility of the surrounding fluids that \(\tsy S^{(1)}_{33}\) and \(\tsy S^{(2)}_{33}\) also have no effect on the flow.
Recall that, for a clean interface, the modes associated with these components of the stresslet (the $S^\perp$ mode) produced a radially symmetric flow associated with local expansion or compression of the interface (see \cref{fig:clean-stresslets}b).
It is no surprise that these source or sink flows vanish for incompressible interfaces.
One may easily verify that there exists no radially symmetric vector field on the interface that both satisfies \(\divS{\vec u} = 0\) and vanishes at infinity.

To further reduce \cref{eq:um1-incompr-expanded}, we break $\vec u^\text{m1}$ into three constituent modes as
\begin{equation}
  \label{eq:um1-by-submode}
  \begin{aligned}
    u^{\text{m1}}_i(\vec x) =\,&
    \lr(){\tsy S^\para_{\ga\gb} + \frac12 L_3 \permut_{3\ga\gb}}
    \Delta^\para_{i\ga\gb}(\vec x) \\ &+
    \frac12 \lr(){\tsy D_{\ga3}^{(1)} + \tsy D^{(2)}_{\ga3}}
    \Delta^\ddagger_{i\ga}(\vec x) +
    \frac12 \lr(){\tsy D_{\ga3}^{(1)} - \tsy D^{(2)}_{\ga3}}
    \Delta^\pm_{i\ga}(\vec x).
  \end{aligned}
\end{equation}
The first mode \(\ten\Delta^\para(\vec x)\) is given by
\begin{equation}
  \label{eq:dipole-special-1}
  \Delta^\para_{i\ga\gb}(\vec x)
  = \lr.|{\frac{\pd \tsy H_{i\ga}(\vec x, \vec y)}{\pd y_\gb}}_{\vec y = \vec 0}
  = -\frac{\pd \tsy H^0_{i\ga}(\vec s, z)}{\pd x_\gb},
\end{equation}
where the second equality follows from the translation invariance of \(\ten H(\vec x, \vec y)\) with respect to components of the singular point $\vec y$ parallel to the interface.
Carrying out the differentiation of \(\ten H^0\) in \cref{eq:dipole-special-1}, we find that
\begin{equation}
  \label{eq:dipole-special-1'}
  \Delta^\para_{\ga\gb\gc}(\vec x) = 
  \frac{R_1'(L_B; s, |z|)}{8\upi \bar\mu} \left(
    \hat s_\ga \kd_{\gb\gc} +
    \hat s_\gb \kd_{\gc\ga} -
    3\hat s_\gc \kd_{\ga\gb}
  \right) -
  \frac{R_3'(L_B; s, |z|)}{2\upi \bar\mu} \Irr{\hat s_\ga \hat s_\gb \hat s_\gc},
\end{equation}
and \(\Delta^\para_{3\gb\gc} = 0\), where the functions \(R'_n\) are given by \cref{eq:rzFn'}.
The first tensorial prefactor of \(\ten\Delta^\para\) in \cref{eq:um1-by-submode} is given by
\begin{equation}
  \tsy S^\para_{\ga\gb} = \left(
    \kd_{\ga\gc} \kd_{\gb\gd} - \frac12 \kd_{\ga\gb} \kd_{\gd\gc}
    \right) \left(
    \tsy S^{(1)}_{\gc\gd} + \tsy S^{(2)}_{\gc\gd} + \tsy S^{(\text i)}_{\gc\gd}
  \right),
\end{equation}
which is the effective stresslet strength in the plane of the interface.

The second mode of \cref{eq:um1-by-submode}, \(\ten\Delta^\ddagger(\vec x)\), describes an equal pair of force dipoles, one just above the interface and the other just below, of strength \(\vec F \vec n\) for $\vec F$ parallel to the interface.
This mode corresponds to the ‘symmetric’ part of the second and third terms of \cref{eq:um1-incompr-expanded} and is given by
\begin{equation}
  \label{eq:dipole-special-2}
  \Delta^\ddagger_{i\ga}(\vec x) =
  \lr(){{\lim_{\vec y \to \vec 0^+}} + {\lim_{\vec y \to \vec 0^-}}}
  \frac{\pd \tsy H_{i\alpha}(\vec x, \vec y)}{\pd h}.
\end{equation}
A convenient property of this mode is that it vanishes on \(z = 0\); differentiation of \cref{eq:Hfun} yields
\begin{equation*}
  \lim_{h \to 0^\pm} \frac{\pd \tsy H^0_{\ga\gb}(\vec s, h)}{\pd h} =
  \pm \lr[]{%
    \frac{R'_0(\lenBq; \vec s, 0)}{4\upi\bar\mu} \kd_{\ga\gb} +
    \frac{R'_2(\lenBq; \vec s, 0)}{2\upi\bar\mu} \lr(){%
      \hat s_\ga \hat s_\gb - \frac12 \kd_{\ga\gb}
    }
  }.
\end{equation*}
The analogous mode for a clean interface \(\ten\Delta^\ddagger_\text{clean}\), given by replacing $\ten H$ with $\ten G$ in \cref{eq:dipole-special-2}, shares the same property.
In fact, we find that
\(
{\lim_{h \to 0} [{\pd \tsy G_{\ga\gb}(\vec x, \vec y)}/{\pd h}]}_{z = 0} = 0.
\)
Therefore, both \(\ten\Delta^\ddagger\) and \(\ten\Delta^\ddagger_\text{clean}\) share the same boundary velocities (which vanish at infinity and on the interface) and external forcing (a pair of equal dipoles on either side of the interface).
Due to uniqueness of solutions to the Stokes equations in each phase, we conclude that
\(\ten\Delta^\ddagger = \ten\Delta^\ddagger_\text{clean}\),
and hence $\ten H$ may be replaced with $\ten G$ in \cref{eq:dipole-special-2} without loss of generality.
We then find from \cref{eq:null-dipole} that
\begin{equation}
  \label{eq:dipole-special-2'}
  \Delta^\ddagger_{i\ga}(\vec x) = -\frac{2}{\mu(z)}
  \left( \kd^\para_{\ga j} n_k + \kd^\para_{\ga k} n_j \right)
  \frac{\pd \tsy J_{ij}(\vec x)}{\pd x_k}
  = \frac{1}{\mu(z)} \left( \frac{3 x_i s_\ga z}{2\upi r^5} \right).
\end{equation}
Comparison of \cref{eq:dipole-special-2'} to the last term of \cref{eq:dipole-moment} shows that \(\ten\Delta^\ddagger\) assumes the same functional form as the ‘$\vec A$ mode’ discussed for clean interfaces in \cref{sec:clean-interfaces}.

The third and final dipolar mode is the antisymmetric compliment of \(\ten\Delta^\ddagger\),
\begin{equation}
  \label{eq:dipole-special-3}
  \Delta^\pm_{i\ga}(\vec x) =
  \lr(){{\lim_{\vec y \to \vec 0^+}} - {\lim_{\vec y \to \vec 0^-}}}
  \frac{\pd \tsy H_{i\alpha}(\vec x, \vec y)}{\pd h},
\end{equation}
which describes a similar pair of force dipoles of opposite sign as they approach the interface from above and below.
To deduce the form of \(\Delta^\pm_{i\ga}\), we use \cref{eq:self-adjoint-H} to reinterpret the tangential stress balance for $\ten H$ \cref{eq:iface-stress-bal-H} as the velocity profile \(\ten W\) due to a pair of force dipoles similar to that represented by \(\ten\Delta^\pm\), except with each dipole weighted by the viscosity of the phase in which it resides.
This velocity profile is given by
\begin{equation}
  \label{eq:iface-stress-bal-H-adjoint}
  \begin{aligned}
    \bar\mu \tsy W_{i\ga}(\vec x) &= \lr(){%
      \mu_1 {\lim_{\vec y \to \vec 0^+}} - {\mu_2 \lim_{\vec y \to \vec 0^-}}
    } \frac{\pd \tsy H_{i\ga}(\vec x, \vec y)}{\pd h} \\ &=
    \lr[]{%
      \frac{\pd \Pi_k(\vec y_s, \vec x)}{\pd y_\beta} -
      \mu_s \frac{\pd^2 \tsy H_{i\ga}(\vec x, \vec y_s)}{\pd y^2_\ga}
    }_{\vec y_s = \vec 0} \\ &=
    \bar\mu \tsy q_{i\ga}(\vec x) - \mu_s \nabla^2 \tsy H^0_{i\ga}(\vec x)
  \end{aligned}
\end{equation}
for \(\vec x \neq \vec 0\), where \(\vec y_s = y_1 \bvec_1 + y_2 \bvec_2\) is the surface projection of $\vec y$ and 
\begin{align}
  \label{eq:surf-p-flow}
  \tsy q_{i\ga}(\vec x) =&\ \frac{1}{\bar\mu} \lr.|{%
    \frac{\pd \Pi_i(\vec y_s, \vec x)}{\pd y_\ga}
  }_{\vec y_s = \vec 0} \\ =&\ 
  \frac{s_i s_\ga}{4\upi\bar\mu} \lr(){%
    \frac{3|z|}{r^5} + \frac{2|z|}{s^4 r} + \frac{|z|}{s^2 r^3} - \frac{2}{s^4}
  } +
  \frac{\kd^\para_{i\ga}}{4\upi\bar\mu} \lr(){%
    -\frac{|z|}{r^3} - \frac{|z|}{s^2 r} + \frac{1}{s^2}
  } - \frac{3 z \lr||{z} s_\ga \delta_{i3}}{4\upi\bar\mu r^5}.
  \nonumber
\end{align}
The mode analogous to \(\ten W\) for a clean interface is null due to continuity of tangential stress \cref{eq:iface-stress-bal-G}.
\Cref{eq:iface-stress-bal-H-adjoint} shows that the gradient of \(\ten\Pi\) can be reinterpreted as the velocity field \(\ten q(\vec x)\) \cref{eq:surf-p-flow}.
This velocity field decays as \(1/r^2\) for all $r$ and is mirror symmetric about $z=0$.
On the \(z=0\) plane, \(\ten q\) takes the same functional form as a two-dimensional source-sink couplet,
\(\tsy q_{\ga\gb}(\vec s) = -\{s_\ga s_\gb\}_0 / 2\pi \bar\mu s^2\).

We may rewrite the first equality of \cref{eq:iface-stress-bal-H-adjoint} as
\begin{equation}
  \label{eq:iface-stress-bal-H-adjoint'}
  \begin{aligned}
    \bar\mu \tsy W_{k\gb}(\vec x) &= \lr[]{ \frac{\mu_1 + \mu_2}{2} \lr(){
        {\lim_{\vec y \to \vec 0^+}} - {\lim_{\vec y \to \vec 0^-}}
        } + \frac{\mu_1 - \mu_2}{2} \lr(){%
        {\lim_{\vec y \to \vec 0^+}} + {\lim_{\vec y \to \vec 0^-}}
      }
    } \frac{\pd \tsy H_{k\gb}(\vec x, \vec y)}{\pd h} \\ &=
    \bar\mu \Delta^\pm_{k\gb}(\vec x) + \frac{\mu_1 - \mu_2}{2} \Delta^\ddagger_{k\gb}(\vec x).
  \end{aligned}
\end{equation}
Solving \cref{eq:iface-stress-bal-H-adjoint'} for \(\ten \Delta^\pm\) and using the final equality of \cref{eq:iface-stress-bal-H-adjoint}, we find that
\begin{equation}
  \label{eq:dipole-special-3'}
  \Delta^\pm_{i\ga}(\vec x) =
  \tsy q_{i\ga}(\vec x) - \lenBq \nabla^2 \tsy H^0_{i\ga}(\vec x) -
  \frac{\mu_1 - \mu_2}{2\bar\mu}\Delta^\ddagger_{i\ga}(\vec x).
\end{equation}
The appearance of \(\nabla^2 \ten H^0\) in \cref{eq:dipole-special-3'} is peculiar because \(\nabla^2 \ten H^0(\vec x) = \nabla^2 \ten H(\vec x, \vec y)|_{\vec y = \vec 0}\) has the functional form of a degenerate quadrupole, whose appearance we anticipate in the quadrupolar term \(\vec u^\text{m2}\) of \cref{eq:mpx-incompr} but not the dipolar term \(\vec u^\text{m1}\).
Indeed, \(\nabla^2 \ten H^0(\vec x)\) decays as \(1/r^3\) for \(r \gg \lenBq\), whereas all other contributions to \(\vec u^\text{m1}\) in \cref{eq:um1-by-submode} decay as \(1/r^2\).
However, \(\ten\Delta^\pm\) describes a mode that is conceptually similar to a ‘usual’ quadrupole: two dipoles of opposite sign separated by a vanishingly small distance.
The key difference here is that these dipoles straddle the interface, creating a Marangoni-driven flow that decays from the singular point as $1/r^2$ in addition to a quadrupolar flow.
This result suggests an adjustment to our original definition of the dipole given by \cref{eq:um1-incompr-expanded} where we exclude the \(\nabla^2 \ten H^0\) term in \(\ten\Delta^\pm\), which is instead incorporated into a modified quadrupolar term (\(\vec u^\text{m2}\)) in \cref{eq:mpx-incompr}.
Denoting the modified dipole as \(\vec {u^\text{m1}}'\), we combine \cref{eq:um1-by-submode,eq:dipole-special-3'} and subtract the degenerate quadrupole from the result to give
\begin{equation}
  \label{eq:um1-by-submode'}
  {u^{\text{m1}}_i}' (\vec x) =
  \lr(){\tsy S^\para_{\ga\gb} + \frac12 L_3 \permut_{3\ga\gb}}
  \Delta^\para_{i\ga\gb}(\lenBq; \vec x) +
  \frac12 A_\ga \Delta^\ddagger_{i\ga}(\vec x) +
  \frac12 B_\ga \tsy q_{i\ga}(\vec x),
\end{equation}
where $\vec A$ is defined by \cref{eq:dipole-coeff-asym}, as it was for a clean interface, and \(B_\ga = (\tsy D^{(1)}_{\ga3} - \tsy D^{(2)}_{\ga3}) \).

\begin{figure}
  \centering
  \includegraphics[width=\linewidth]{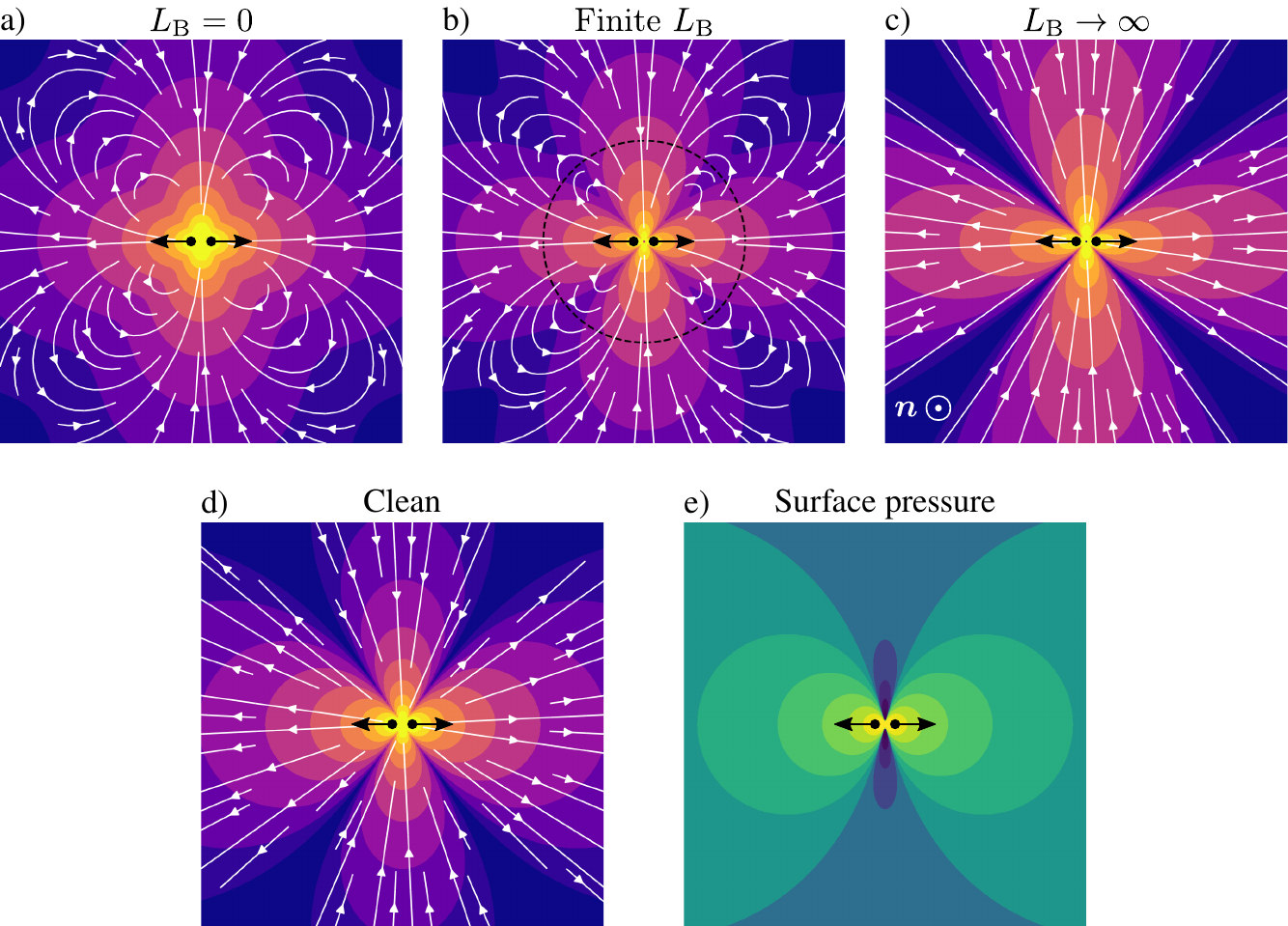}
  \caption{%
    The interfacial velocity and surface pressure fields associated with the $\ten S^\para$ mode are shown for the case of an extensile force dipole.
    The panels are analogous to those shown in \cref{fig:monopoles}.
    The arrows indicate the orientation of the force dipole.
    Panels (a), (b), and (c) represent one of three leading-order flow modes induced by self-propelled (force- and torque-free) colloids for small-, moderate-, and large-\(\numBq\), respectively.
  }%
  \label{fig:dipoles}
\end{figure}

Each term of \cref{eq:um1-by-submode'} represents a distinct contribution to \({\vec u^\text{m1}}'\).
Only the first term of \({\vec u^\text{m1}}'\) in \cref{eq:um1-by-submode'} is sensitive to surface viscosity, hence our inclusion of $\lenBq$ as a argument to \(\ten\Delta^\para\) here.
Like the monopolar velocity field, \(\ten\Delta^\para\) produces a ‘lamellar’ fluid motion, with no flow normal to the interface.
Its coefficients \(\ten S^\para\) and \(L_3\) are the tensorial strengths of the stresslet parallel to the interface and the rotlet normal to the interface, respectively.
The associated modes differ significantly from their counterparts on a clean interface.
The $\ten S^\para$ (stresslet) mode is affected by Marangoni and surface-viscous stresses; it is distinct from the analogous mode on a clean interface even if \(\lenBq = 0\).
\Cref{fig:dipoles} visualizes the \(\ten S^\para\) contribution to \({\vec u^\text{m1}}'\) for different values of \(\lenBq\).
On the other hand, the \(L_3\) (rotlet) mode does not generate a surface pressure gradient because \(\ten\Pi(\vec s, h=0)\) is harmonic.
It is therefore affected by surface viscosity but does not contribute to Marangoni stresses.
For an interface with vanishing surface viscosity, the $L_3$ mode for an incompressible interface does not differ from that of a clean interface; both assume the functional form of a rotlet in a bulk fluid of viscosity \(\bar\mu\).

Examining the behavior of \(\ten\Delta^\para\) for \(\lenBq = 0\), \cref{eq:dipole-special-1',eq:rzFn'|Bq=0} give
\begin{equation}
  \label{eq:dipole-special-1|Bq=0}
  \lr.|{\Delta^\para_{\ga\gb\gc}(\vec x)}_{\lenBq=0} = 
  \frac{(3r + |z|) \Irr{s_\ga s_\gb s_\gc}}{4\upi \bar\mu r^3 (r + |z|)^3} -
  \frac{s_\ga \kd_{\gb\gc} + s_\gb \kd_{\gc\ga} - 3 s_\gc \kd_{\ga\gb}}
       {16 \upi \bar\mu r^3},
\end{equation}
which decays spatially as \(1/r^2\), similar to the analogous $\ten S^\para$ and $L_3$ modes for a clean interface.
As \(\lenBq \to \infty\), \(\ten\Delta^\para\) approaches the functional form of a two-dimensional Stokeslet dipole, given by the gradient of equation \cref{eq:Hfun0|Bq=oo},
\begin{equation}
  \label{eq:Stokeslet-dipole-2d}
  \newcommand*\hs{\hat s}
  \Delta^\para_{\ga\gb\gc}|_{\lenBq\to\infty} \sim
  \frac{%
    2 \hs_\ga \hs_\gb \hs_\gc
    - \hs_\ga \kd_{\gb\gc} - \hs_\gb \kd_{\gc\ga} + \hs_\gc \kd_{\ga\gb}
  }{4\upi\mu_\surf s}.
\end{equation}
Thus, this dipolar mode decays as \(1/s\) as \(s \to \infty\), as opposed to the logarithmically divergent behavior of the monopole in the same limit.
However, as a three-dimensional flow, \cref{eq:Stokeslet-dipole-2d} is constant along the $z$ axis, in conflict with the condition that $\vec u$ vanishes as \(z \to \infty\).
This behavior is apparently another manifestation of Stokes' paradox.
Physically, if we consider a colloid of size $a$ and \(\numBq\) large but finite, bulk-viscous effects eventually dominate over surface-viscous effects at distances \(r/a \gg \numBq\) from the colloid, and a \(1/r^2\) velocity decay is recovered in the far field.

The second term in \cref{eq:um1-by-submode'}, the $\vec A$ mode, is unaltered from the expression for \(\vec u^\text{m1}\) for a clean interface \cref{eq:dipole-moment}.
Of the three terms, only this one is sensitive to the viscosity contrast.
We refer the reader back to \cref{fig:clean-stresslets}c for a depiction of the associated velocity field.
Marangoni and surface-viscous stresses have no effect on this mode, which fails to induce motion of the interface.
Its interpretation regarding colloid asymmetry is therefore the same as discussed in \cref{sec:clean-interfaces}.

\begin{figure}
  \centering
  \includegraphics[width=\linewidth]{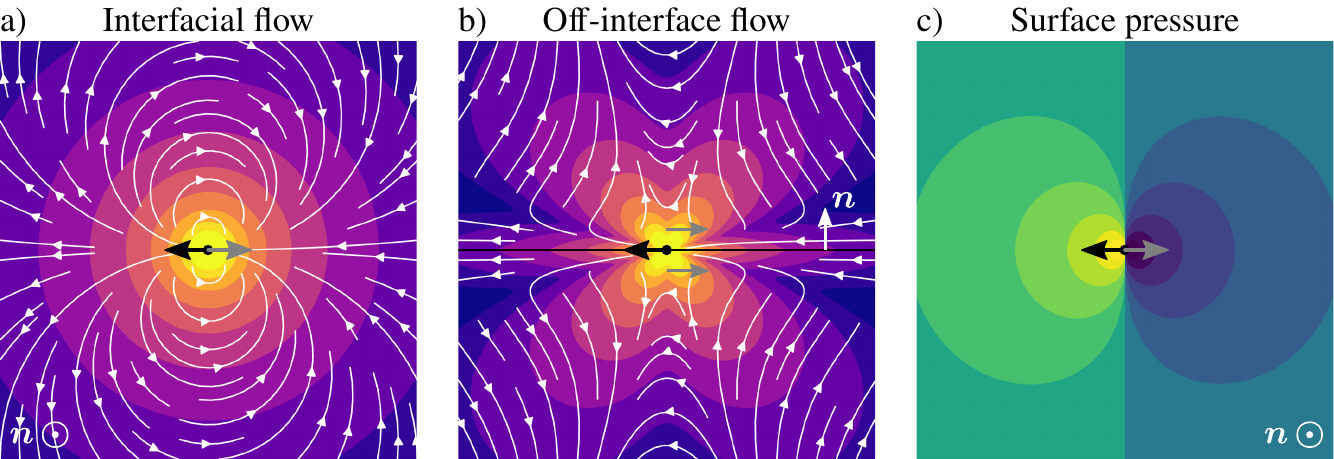}
  \caption{%
    The velocity field \(\vec B \vdot \ten q(\vec x)\) is visualized for the vector $\vec B$ pointing rightward.
    This mode is associated with a (viscosity-weighted) pair of force couplets indicated by the arrows.
    The flow on the interface is shown in (a), and a perpendicular cross-section of the off-interface flow is shown in (b).
    The associated surface pressure is shown in (c).
  }%
  \label{fig:dipole-q}
\end{figure}

The final term of \cref{eq:um1-by-submode}, the $\vec B$ mode, has no analogue on a clean interface.
Physically, the $\vec B$ mode accounts for the coupling of off-interface forcing of the fluid by the colloid to Marangoni forces that react to enforce interfacial incompressiblity.
Interestingly, the coefficient $\vec B$ does not explicitly involve forces on the contact line, in contrast to $\ten S^\para$ and $L_3$, but the resulting flow is nonetheless driven by the surface pressure according to \cref{eq:surf-p-flow}.
The interfacial flow field due to this mode is harmonic, \(\nabla_\surf^2 \ten q(\vec s) = \vec 0\) (\cref{fig:dipole-q}a).
By \cref{eq:iface-stress-bal-Newtonian}, it therefore does not induce surface-viscous stresses, explaining why $\ten q$ is independent of $\lenBq$.
Like the $\ten S^\para$ and $L_3$ modes, the $\vec B$ mode produces a mirror-symmetric velocity field about the interface.
However, while the $\ten S^\para$ and $L_3$ modes produce a lamellar velocity field, the velocity field due to the $\vec B$ mode is fully three-dimensional (\cref{fig:dipole-q}b).
The surface pressure associated with $\tsy q_{i\ga}(\vec x)$ is given by \({\pd \Pi_\ga(\vec s, h \bvec_3)} / {\pd h} = - s_\ga / 2\pi s^3\) (\cref{fig:dipole-q}c).

Both the $\vec B$ and $\vec A$ modes can be attributed to the finite protrusion of an adhered colloid from the interface.
As the protrusion distance becomes small (or as the geometry of the colloid becomes flat) $\vec A$ and $\vec B$ vanish.
It is interesting that finite-size effects enter at dipolar order in this manner.
Consider again the velocity profile due to a sphere straddling a clean interface with a \SI{90}{\degree} contact angle.
Here, the size of the sphere matters only at quadrupolar order, through the \(a^2 \nabla^2 \ten J\) term in \cref{eq:sphere-flow}.
In contrast to the \(1/r^3\) decay of this term, the dipolar $\vec A$ and $\vec B$ modes decay as $1/r^2$.
On an incompressible (or clean) interface, $\vec A$ vanishes by symmetry for this particular case, but we expect \(\vec B \sim a \vec F\), where $\vec F$ is the force driving the colloid.
It is intriguing how this finite-protrusion effect arises from interfacial Marangoni forces.

\subsection{Discussion}

Interfacial incompressibility dramatically restructures the leading-order hydrodynamic modes induced by colloidal particles at interfaces.
Marangoni stresses play a major role; even without surface viscosity, only the $L_3$ (rotlet) and $\vec A$ modes are unchanged from their counterparts on a clean interface.
Marangoni forces also lead to a mode of flow, the $\vec B$ mode given by the final term of \cref{eq:um1-by-submode'}, that has no analogue on a clean interface.

Recall that, at clean interfaces, the far-field fluid velocity decays at the same rate in directions parallel and normal to the interface: generally, \(|\vec u| \sim r^{-1}\) for driven colloids and \(|\vec u| \sim r^{-2}\) for active colloids (or colloids driven only by an external torque).
If surface-viscous stresses are weak, \(\numBq \ll 1\), then this far-field behavior also holds for incompressible interfaces.
However, for driven colloids, the leading-order flow normal to the interface is hindered due to the lamellar profile of the force monopole \cref{eq:monopole-moment-incompr}.
For driven and active colloids alike, the fluid velocity normal to the interface is generally induced to leading order by the dipolar $\vec A$ and $\vec B$ modes (that decay as $1/r^2$).
The $\vec A$ and $\vec B$ modes therefore have special importance to mass transport enhancement and hydrodynamic interactions normal to the interface.
Interestingly, these modes are unaffected by surface viscosity.

However, surface-viscous effects can give rise to very long-ranged flow parallel to the interface.
Considering first the spatial behavior along the interfacial plane, we find that, for \(\numBq \gg 1\) and distances \(s \ll \lenBq\) from the colloid, \(\vec u^\para = \vec u \vdot \ten I_\surf \sim -\ln s \) for the monopole moment and \(\vec u^\para \sim s^{-1}\) for dipole moment.
This behavior reflects that of a two-dimensional Stokes flow, which is recovered in the limit of highly viscous interfaces \citep{Saffman1975}.
The slow decay of the velocity field quickens at distances \(s \gtrsim \lenBq\), where bulk-viscous effects inevitably become important.
To determine the spatial behavior along the \(z\) axis, we observe that the limiting forms of \cref{eq:rzFn,eq:rzFn'} for \(s \ll |z|\) are given by
\begin{align}
  R_n(\lenBq; 0, |z|) &\sim
  \frac{e^{2|z|/\lenBq}}{\lenBq} \ExpInt_1 \!\lr(){\frac{2|z|}{\lenBq}}
\\
  R'_n(\lenBq; 0, |z|) &\sim
  \frac{e^{2|z|/\lenBq}}{\lenBq\lr||z} \ExpInt_2 \!\lr(){\frac{2|z|}{\lenBq}},
\end{align}
where \(\ExpInt_p (x) = \int_1^\infty e^{-xt} / t^p \dd t\) is the generalized exponential integral.
The asymptotic behavior of this function obeys
\begin{equation}
  \ExpInt_p(x) \sim
  \begin{cases}
    [{(-1)}^p / (p-1)!] x^{p-1} \ln x     & \text{for}\ x \ll 1 \\
    \ExpInt_p(x) \sim e^{-x}/x & \text{for}\ x \gg 1\ \text{and for all}\ p,
  \end{cases}
\end{equation}
\citep{Olver2010} which implies that, for \(|z| \ll \lenBq\), \(R_n \sim \ln |z|\) and \(R'_n \sim \ln |z|\).
Recalling that \(R_n\) and \(R'_n\) govern the spatial behavior of the monopole and $\ten S^\para$ (dipole) modes, respectively, we see that both are logarithmically divergent in $z$ as $\numBq$ is made large.
Therefore, the ‘lamellar’ flows these modes induce persist up to distances \(z = O(\lenBq)\) into the surrounding fluid for both driven and active colloids.
At distances \(z \gg \lenBq\), we find that \(R_n \sim |z|^{-1}\) and \(R'_n \sim |z|^{-2}\), recovering the far-field decay expected for \(r \gg \lenBq\).
This strong lateral fluid motion has potential implications for mixing generation by colloidal particles at interfaces.
For instance, in the presence of a driven or active sheet of interfacially trapped colloids, there may be significant mass transport in directions parallel to the interface via Taylor dispersion.
The shear flow driving Taylor dispersion is, in this case, generated by the motion of the colloids rather than motion of a bulk fluid relative to a no-slip boundary.
Future work will examine the implications of this fluid motion on transport and mixing rates at interfaces with colloids present.

\section{Conclusion}
\label{sec:conclusion}

\subsection{Summary}

We have determined the leading-order far-field flows generated by driven and active colloids trapped at planar fluid interfaces by a pinned contact line for $\numCa \ll 1$.
Under these assumptions, the colloid is trapped in a fixed wetting configuration and cannot move perpendicular to the interface.
At clean interfaces devoid of surfactant, driven colloids produce “viscosity-averaged” Stokeslets when driven along the interface---the flow is no different than that expected for a colloid driven in an unbounded fluid of viscosity $\bar\mu$.
Contact-line pinning at small $\numCa$ prevents such colloids from being driven normal to the interface.
Similarly, active colloids produce viscosity-averaged force dipoles (stresslets) aligned in the swimming direction, similar to those generated by a swimmer in an unbounded fluid.
This stresslet is associated with balanced hydrodynamic thrust and drag in the swimming direction.
However, such swimmers also generate additional ‘pumping’ flows that are associated with net hydrodynamic forces and torques on the colloid that are supported by the pinned contact line.
These modes vanish if the colloid is adjacent, rather than adhered, to the interface, where it must be force and torque free.

We further consider the effect of surfactants, which render the interface incompressible even in the limit of scant surface concentrations.
This constraint is generally applicable to driven and active colloids which move on interfaces for $\numCa \ll 1$.
The flow modes associated with forced or self-propelled motion along the interface are altered significantly by interfacial incompressibility.
One subset of these modes induce ‘lamellar’ fluid motion for which $u_z = 0$ at all distances $z$ from the interface.
These modes can produce very long ranged flow in the presence of surface-viscous effects.
A second subset of modes is responsible for the leading-order fluid motion normal to the interface.
These modes occur at dipolar order and are unaffected by surface viscosity.

\subsection{Future work and open issues}

Future work will probe experimental systems for signatures of the flow modes reported here.
For example, the differences we predict for colloids in adhered states versus unadhered states may be useful in distinguishing between these two cases in experiment.
Comparison to computational results would also be extremely valuable.
Detailed numerical computations of specific types of colloids in different adhered states will yield useful information such as their predicted trajectories and near-field hydrodynamic interactions.

Several open issues remain.
We have not considered the effect of contact-line undulations on the flow.
Interestingly, interfacial distortion due to such undulations spatially decay at the same rate ($1/r^2$) as the flow disturbance due to an active colloid of negligible weight.
Thus, these undulations may alter the flows in interesting ways, especially because the contact line of an individual colloid may undulate randomly, being different for every colloid \citep{Stamou2000,Kaz2012}.
Driven and active colloids may also enhance mass transport at interfaces.
Enhanced mixing in active colloidal suspensions has been studied extensively in bulk fluids \citep{Darnton2004,Pushkin2013,Lin2011,Kasyap2014} and also in the vicinity of solid boundaries \citep{Mathijssen2015,Mathijssen2018,Kim2004}.
Colloid-induced mixing presents an untapped dimension for interfacial engineering; interfaces are natural sites for many chemical reaction and separation processes.
At interfaces, mixing rates will depend on the interfacial rheology and the adhered state of the active colloids that populate the interface.
Our work emphasizes the importance of particular far-field flow modes in the generation of mixing by active or passive colloids at interfaces.
For an incompressible interface, a subset of these modes promote fluid motion parallel to the interface, which is especially long-ranged for viscous interfaces, and another subset promotes fluid motion normal to the interface.
Future work will determine design guidelines for enhancing or directing mass transport at interfaces via colloidal particles by tuning the relative strengths of these modes.

The authors acknowledge useful discussions with Dr.\ Mehdi Molaei and Ms.\ Jiayi Deng.
This research was made possible by a grant from the Gulf of Mexico Research Initiative.
Additional financial support was provided by the National Science Foundation (NSF Grant No.\ DMR-1607878 and CBET-1943394).

Declaration of Interests. The authors report no conflict of interest.

\appendix

\section{Self-adjoint property of the Green's functions}
\label{sec:app:self-adjointness}

To show that the Green's function \(\ten G\) defined by \cref{eq:Gfun-clean} is self-adjoint, i.e., \(\ten G(\vec x, \vec y) = \ten G^\T(\vec y, \vec x)\), we make the following substitutions into \cref{eq:R-thm-clean-iface}:
\begin{equation}
  \label{eq:sa-subs}
  \begin{aligned}
    \vec u(\vec x)      &\to \ten G(\vec x, \vec y) \vdot \vec F,           &%
    \vec u'(\vec x)     &\to \ten G(\vec x, \vec y') \vdot \vec F',         \\
    \ten\sigma(\vec x)  &\to \ten T(\ten G; \vec x, \vec y) \vdot \vec F,   &%
    \ten\sigma'(\vec x) &\to \ten T(\ten G; \vec x, \vec y') \vdot \vec F', \\
    \vec f(\vec x)      &\to -\vec F \diracR3(\vec x - \vec y),             &%
    \vec f'(\vec x)     &\to -\vec F \diracR3(\vec x - \vec y'),            \\
    \vec f_\surf(\vec x)    &\to -\vec F \diracR2(\vec x - \vec y),             &%
    \vec f_\surf'(\vec x)   &\to -\vec F \diracR2(\vec x - \vec y').
  \end{aligned}
\end{equation}
That is, we choose \(\vec u\) as the flow field due to a point force \(\vec F\) at \(\vec y\) and \(\vec u'\) the flow field due to another point force \(\vec F'\) at point \(\vec y'\).
The point forces and their locations are arbitrary and may be exerted on either fluid or the interface.
Each fluid domain is semi-infinite and bounded only by the interface.
With the above substitutions, \cref{eq:R-thm-clean-iface} becomes
\begin{multline}
  \label{eq:sa-intermediate}
  0 =
  \int_{V^*} \left[
    \diracR3(\vec x - \vec y) \vec F \vdot
      \ten G(\vec x, \vec y') \vdot \vec F' - 
    \diracR3(\vec x - \vec y') \vec F' \vdot
      \ten G(\vec x, \vec y) \vdot \vec F
  \right] \dd V \\ +
  \int_{I^*} \left[
    \diracR2(\vec x - \vec y) \vec F \vdot
      \ten G(\vec x, \vec y') \vdot \vec F' - 
    \diracR2(\vec x - \vec y') \vec F' \vdot
      \ten G(\vec x, \vec y) \vdot \vec F
  \right] \dd A \\ +
  \oint_{R} \left\{
    [\ten T(\vec x, \vec y) \vdot \vec F] \vdot
      [\ten G(\vec x, \vec y') \vdot \vec F'] -
    [\ten T(\vec x, \vec y') \vdot \vec F'] \vdot
      [\ten G(\vec x, \vec y) \vdot \vec F]
  \right\} \vdot \uvec n \dd S,
\end{multline}
where, for brevity, we omit \(\ten G\) as an argument to \(\ten T\).
The integrations in \cref{eq:sa-intermediate} are taken to be over an arbitrary volume that may contain points on the interface.
If the boundaries of this volume in each fluid, represented by \(R\), are made arbitrarily far from the points \(\vec y\) and \(\vec y'\), then the final integral in \cref{eq:sa-intermediate} vanishes; \(\ten G \sim r^{-1}\) and \(\ten T(\ten G;\!) \sim 1/r^{-2}\), so this integral decays as \(L_V^{-1}\) as \(L_V \to \infty\), where \(L_V\) is the characteristic size of the integration region.
Then, using the definition of the Dirac delta, \cref{eq:sa-intermediate} simplifies to
\begin{equation}
  \label{eq:G-is-self-adjoint}
  \vec F \vdot \ten G(\vec y, \vec y') \vdot \vec F' =
  \vec F' \vdot \ten G(\vec y', \vec y) \vdot \vec F.
\end{equation}
Since \(\vec F\), \(\vec F'\), \(\vec y\), and \(\vec y'\) are all arbitrary, \cref{eq:G-is-self-adjoint} implies that \(\ten G(\vec y, \vec y') = \ten G^\T(\vec y', \vec y)\), that is, \(\ten G\) is self-adjoint.

Using the same procedure, it may be shown that \(\ten H\) is also self-adjoint.
Making a set of substitutions analogous to those appearing in \cref{eq:sa-subs} along with the additional substitutions \(\ten\varsigma(\vec x) \to \ten\Sigma(\ten H; \vec x, \vec y) \vdot \vec F\) and \(\ten\varsigma(\vec x) \to \ten\Sigma(\ten H; \vec x, \vec y') \vdot \vec F'\) into \cref{eq:R-thm-incompressible-iface}, we find
\begin{multline}
  \label{eq:sa-intermediate-H}
  0 =
  \vec F' \vdot \ten H(\vec y', \vec y) \vdot \vec F -
  \vec F \vdot \ten H(\vec y, \vec y') \vdot \vec F' \\ +
  \oint_{R} \left\{
    [\ten T(\vec x, \vec y) \vdot \vec F] \vdot
      [\ten H(\vec x, \vec y') \vdot \vec F'] -
    [\ten T(\vec x, \vec y') \vdot \vec F'] \vdot
      [\ten H(\vec x, \vec y) \vdot \vec F]
  \right\} \vdot \uvec n \dd S \\ +
  \oint_{\partial I^*} \left\{
    [\ten\Sigma(\vec x, \vec y) \vdot \vec F] \vdot
      [\ten H(\vec x, \vec y') \vdot \vec F'] -
    [\ten\Sigma(\vec x, \vec y') \vdot \vec F'] \vdot
      [\ten H(\vec x, \vec y) \vdot \vec F]
  \right\} \vdot \uvec m \dd C.
\end{multline}
In this case, an additional integral over \(\partial I^*\) appears, which is the curve where our arbitrarily chosen fluid region intersects the interface.
Both integrals in \cref{eq:sa-intermediate-H} vanish as \(L_V \to \infty\) provided that the Boussinesq length \(\lenBq\) remains finite.
For \(r \gg \lenBq\), \(\ten H\) and \(\ten G\) share the same far-field decay behavior, i.e., \(\ten H \sim r^{-1}\).
From the remaining two terms in \cref{eq:sa-intermediate-H}, we find \(\ten H(\vec y, \vec y') = \ten H^\T(\vec y', \vec y)\).

\section{Derivation of the Green's functions}
\label{sec:app:Gfuns}

Here, we derive the Green's functions \(\ten G\) and \(\ten H\) used in \cref{sec:clean-interfaces,sec:incompressible-interfaces}, respectively.
We consider the scenario described in \cref{sec:reciprocal-relations}, where two immiscible fluids are separated by a flat interface on the $z=0$ plane, except with no particles present.
The velocity field in a region $V^*_\nu$ that is fully contained in fluid $\nu$ can be represented in boundary integral form as \citep{Kim1991,Pozrikidis1992}
\begin{multline}
  \label{eq:BIE-fundamental}
    \Ind_{V^*_\nu}(\vec x) \, \vec u(\vec x) =
    - \frac{1}{\mu_\nu} \oint_{\partial V^*_\nu} {%
      \ten J(\vec x, \vec y)
      \vdot [\ten\sigma \vdot \uvec n ](\vec y)
    } \dd S(\vec y)
    + \oint_{\partial V^*_\nu} {%
      {[\vec u \uvec n]}(\vec y) \odot \ten T (\ten J; \vec y, \vec x)
    } \dd S(\vec y) \\ %
    + \frac{1}{\mu_\nu} \int_{V^*_\nu} {%
      \ten J(\vec x, \vec y) \vdot \vec f(\vec y)
    } \dd{V(\vec y)},
\end{multline}
where
\(\tsy J_{ij}(\vec x) = (\kd_{ij}/|\vec x| + x_i x_j / {|\vec x|}^3)/8\upi\),
\(\tsy T_{ijk}(\ten J;\!) = -\delta_{ij} P_k(\ten J;\!) + (\nabla_i \tsy J_{jk} - \nabla_j \tsy J_{ik})\),
$\uvec n$ is the inward-facing unit normal of \(\partial V_\nu^*\), and $\vec f$ is the force density on the fluid.
For notational convenience, we hereafter omit \(\ten J\) as an argument to \(\ten T\).

We choose a point \(\vec x = \vec y = h \bvec_3\) at which a point force $\vec F$ is applied.
Thus, we set \(\vec f(\vec y) = \vec F \diracR3(\vec x - \vec y)\).
We then apply \cref{eq:BIE-fundamental} to a volume $V^*_1$ in fluid 1 whose boundary $\partial V^*_1$ is on one side completely adjacent to the interface $I$ and at all other points is made arbitrarily far from the point of forcing $\vec y$.
We repeat this process for a similar volume $V^*_2$ in fluid 2 such that \(V^*_\nu \to V_\nu\) and \(\partial V^*_1 \cup \partial V^*_2 \to I\).
In the resulting pair of equations, only integrations over $I$ make a non-vanishing contribution to $\vec u(\vec x)$ in \cref{eq:BIE-fundamental}.
Fourier transformation of these equations gives
\begin{align}
  \label{eq:FT-upper-fluid}
  \Ind_{\Reals_+}\!(z) \, \FT u_i(\vec k, z) &=
    - \FT{\mathsfi T}_{3 \ga i}(\ten J; \vec k, z) \FT v_\ga(\vec k)
    - \frac{1}{\mu_1} \FT{\mathsfi J}_{ij}(\vec k, z) \FT t^1_j(\vec k)
    + \frac{\Ind_{\Reals_+}\!(h)}{\mu_1} \FT{\mathsfi J}_{ij}(\vec k, z-h) F_j \\
  \label{eq:FT-lower-fluid}
  \Ind_{\Reals_-}\!(z) \, \FT u_i(\vec k, z) &=
    \FT{\mathsfi T}_{3 \ga i}(\ten J; \vec k, z) \FT v_\ga(\vec k)
    + \frac{1}{\mu_2} \FT{\mathsfi J}_{ij}(\vec k, z) \FT t^2_j(\vec k),
    + \frac{\Ind_{\Reals_-}\!(h)}{\mu_2} \FT{\mathsfi J}_{ij}(\vec k, z-h) F_j,
\end{align}
where the Fourier transform is defined as
\(\FT \phi(\vec k) := \iint_{\Reals^2} \phi(\vec s)
  \exp{(-i \vec k \vdot \vec s)} \dd^2 \vec s\),
with \(\vec s = x_1 \bvec_1 + x_2 \bvec_2\) denoting the position vector on the interface.
In \cref{eq:FT-upper-fluid,eq:FT-lower-fluid}, \(\vec t^\nu = \bvec_3 \vdot \vec\sigma^\nu|_{z=0}\) is the surface traction on the fluid-$\nu$-side of the interface, and \(\vec v(\vec s) = \vec u(\vec s, z=0)\) is the surface velocity on the interface.
The Fourier transform of $\ten J$ is given by
\begin{equation}
  \label{eq:FT-Stokeslet}
  \FT{\mathsfi J}_{ij}(\vec k, z) = \frac{\delta_{ij}}{2k} e^{-k|z|}
  + \frac{1}{4k^3} \FT{\nabla}_i \FT{\nabla}_j \left[
    (1 + k|z|) e^{-k|z|}
  \right].
\end{equation}
where $\FT{\nabla}_i := ik_i + \delta_{i3} (\partial/\partial z$).

From \cref{eq:iface-stress-bal-clean}, the Fourier transform of the tangential stress balance on a clean interface is given by
\begin{equation}
  \label{eq:FT-interface-clean}
  \ten I_\surf \vdot \bigl( {[\FT{\vec t}]}_I + \Ind_{\{0\}}(h) \, \vec F \bigr) = \vec 0,
\end{equation}
and for an incompressible interface from \cref{eq:iface-stress-bal-Newtonian} by
\begin{equation}
  \label{eq:FT-interface-incompr}
  \ten I_\surf \vdot \bigl( {[\FT{\vec t}]}_I + \Ind_{\{0\}}(h) \, \vec F \bigr) = i \vec k \FT\pi + \mu_\surf k^2 \FT{\vec v}; \quad
  i\vec k \vdot \vec v = 0,
\end{equation}
where $k = |\vec k|$.

We multiply \cref{eq:FT-upper-fluid} by $\mu_1$ and take the limit of the resulting equation as \(z \to 0^+\).
Similarly, we multiply \cref{eq:FT-lower-fluid} by $\mu_2$ and take the limit \(z \to 0^-\).
Adding these two results, we find
\begin{equation}
  \label{eq:Delta-force-balance-full}
  2\bar\mu \kd_{i\gb} \FT v_\gb(\vec k)
  + \left( \mu_1 \lim_{z\to0^+} - \mu_2 \lim_{z\to0^-} \right)
    \FT{\mathsfi T}_{3 \ga i}(\vec k, z) \FT v_\ga (\vec k)
    + \FT{\mathsfi J}_{ij}(\vec k, 0) \lr[]{\FT t_j}_I (\vec k)
  = \FT{\mathsfi J}_{ij}(\vec k, -h) F_j,
\end{equation}
where $\bar\mu = (\mu_1 + \mu_2) / 2$ is the average viscosity.
Using \cref{eq:FT-Stokeslet} and the definition of $\ten T$, we find that the second term on the left-hand side of \cref{eq:Delta-force-balance-full} reduces to
\begin{equation}
  \label{eq:FT-double-layer-jump}
  \lim_{z\to0^\pm} \FT{\mathsfi T}_{3 \ga i} \FT v_\ga
  = - \left( \pm\frac{\delta_{i\ga}}{2} + \delta_{i3} \frac{ik_\ga}{2k} \right)
    \FT v_\ga,
\end{equation}
which is the Stokes ‘double-layer’ density for either side of the interface.
For a clean interface, \cref{eq:FT-interface-clean,eq:FT-double-layer-jump} in \cref{eq:Delta-force-balance-full} yields, after a trivial Fourier inversion,
\begin{equation}
  \label{eq:FT-vel-surf-clean}
  \bar\mu \vec v(\vec s) = \ten I_\surf \vdot \ten J(\vec s - h \bvec_3) \vdot \vec F,
\end{equation}
which shows that the fluid velocity at the interface is independent of the viscosity contrast and simply corresponds to the projection of $\ten J$, shifted to $z=h$, onto the interface at $z=0$.

We may do the same for an incompressible interface by instead using \cref{eq:FT-interface-incompr} in \cref{eq:Delta-force-balance-full}, from which we obtain
\begin{equation}
  \label{eq:Delta-force-balance}
  \left(\bar\mu + \frac12 \mu_\surf k\right) \FT v_\ga + \frac{ik_\ga}{4k} \FT\pi
  = \FT{\mathsfi J}_{\ga j}|_{z=-h} F_j.
\end{equation}
Taking the inner product of \cref{eq:Delta-force-balance} with $i\vec k$ and solving for $\FT\pi$ yields
\begin{equation}
  \label{eq:FT-surface-pressure}
  \begin{aligned}
    \FT\pi(\vec k)
    &= -\frac{4}{k} i\vec k \vdot \FT{\mathsfbi J}(\vec k, -h) \vdot \vec F \\
    &= \frac{e^{-k |h|}}{k^2} \left[
      (k |h| - 1) i\vec k + k^2 h \vec i_z
    \right] \vdot \vec F.
  \end{aligned}
\end{equation}
Letting $\pi(\vec s, h) = \vec\Pi(\vec s, h) \vdot \vec F$ and carrying out the Fourier inversion to real space, we obtain
\begin{equation}
  \label{eq:surface-pressure'}
  \newcommand*\rootrq{\sqrt{s^2 + h^2}}
  \vec\Pi(\vec s, h)
  = |h| \left( \gradS - \bvec_3 \frac{\pd}{\pd h} \right) \frac{1}{4\upi\rootrq}
  + \frac{\vec s}{4\upi s^2} \left( 1 - \frac{|h|}{\rootrq} \right),
\end{equation}
which reduces to equation \cref{eq:surface-pressure} of the main text.

Noting that \(\vec v(\vec s) \equiv \ten H(\vec x, \vec y)|_{z=0} \vdot \vec F\), we combine \cref{eq:FT-surface-pressure} and \cref{eq:Delta-force-balance} and solve for $\FT{\vec v}$ to give
\begin{equation}
  \label{eq:FT-vel-surf-incompr}
  \begin{aligned}
    \FT{\mathsfbi H} (\vec k, z=0, h) &= \frac{2}{2\bar\mu + \mu_\surf k} \left(
      \ten I_\surf - \frac{\vec{kk}}{k^2}
    \right) \vdot \FT{\mathsfbi J}(\vec k, -h)
    \\
    &= \frac{e^{-k |h|}}{2\bar\mu + \mu_\surf k} \left(
      \frac{\ten I_\surf}{k} - \frac{\vec{kk}}{k^3}
    \right).
  \end{aligned}
\end{equation}
Surface incompressibility of $\ten H$ is easily verified by contracting the right-hand side of \cref{eq:FT-vel-surf-incompr} with \(i\vec k\), thereby taking the divergence in Fourier space, which vanishes.
We also see from \cref{eq:FT-vel-surf-incompr} that a force perpendicular to the interface generates no interfacial flow; \(\tsy H_{i3}(\vec x \in I, h) = 0\).
We conclude that the flow due to the \(z\)-component of the force is the same as that for a rigid, no-slip wall, as is also noted by \citet{Blawzdziewicz1999}.

The self-adjoint property of \(\ten H\) \cref{eq:self-adjoint-H} permits us to swap the roles of \(h\) and \(z\) in \cref{eq:FT-vel-surf-incompr};
\begin{equation}
  \label{eq:FT-Hfun}
  \FT{\mathsfbi H} (\vec k, z, h=0) :=
  \FT{\mathsfbi H}^0(\vec k, z) = \frac{e^{-k |z|}}{2\bar\mu + \mu_\surf k} \left(
    \frac{\ten I_\surf}{k} - \frac{\vec{kk}}{k^3}
  \right).
\end{equation}
From the interfacial flow profile due to a point force at \(z = h\) \cref{eq:FT-vel-surf-incompr}, we automatically obtain the flow at all points \(\vec x\) due to a point force at the interface (\(h = 0\)).
Fourier inversion of \cref{eq:FT-Hfun} to real space gives
\begin{equation}
  \label{eq:app:Hfun} 
  \tsy H^0_{\ga\gb}(\lenBq; \vec s, z) = 
  \frac{1}{4\upi \bar\mu} R_0(\lenBq; s, |z|) \kd_{\ga\gb} +
  \frac{1}{2\upi \bar\mu} R_2(\lenBq; s, |z|) \Irr{\hat s_\ga \hat s_\gb},
\end{equation}
where \(\lenBq = \mu_\surf / \bar\mu \) and \(s = |\vec s|\).
\Cref{eq:app:Hfun} is the same as \cref{eq:Hfun} of the main text.
The functions \(R_n\) are given by
\begin{equation}
  \label{eq:rzFn}
  R_n(\lenBq; s, \zeta) =
  \int_0^\infty \frac{e^{-k \zeta}}{2 + \lenBq k} J_n(ks) \dd k,
\end{equation}
where \(J_n\) is the Bessel function of the first kind of order \(n\).
For \(\lenBq = 0\), \(R_n\) reduces to
\begin{equation}
  \label{eq:rzFn|Bq=0}
  R_n(0; s, |z|) = \frac{(r - |z|)^n}{2r s^n},
\end{equation}
where \(r = \sqrt{s^2 + z^2}\).
To obtain (surface) gradients of \cref{eq:app:Hfun}, we may take the tensor product of \cref{eq:FT-Hfun} with \(i\vec k\) and repeat the Fourier inversion process to give
\begin{equation}
  \frac{\pd \tsy H^0_{\ga\gb}}{\pd s_\gc} =
  -\frac{R_1'(\lenBq; s, |z|)}{8\upi \bar\mu} \left(
    \hat s_\ga \kd_{\gb\gc} + \hat s_\gb \kd_{\gc\ga} -
    3\hat s_\gc \kd_{\ga\gb}
  \right) +
  \frac{R_3'(\lenBq; s, |z|)}{2\upi \bar\mu} \Irr{\hat s_\ga \hat s_\gb \hat s_\gc},
\end{equation}
where
\begin{equation}
  \label{eq:rzFn'}
  R'_n(\lenBq; s, \zeta) = \frac{\pd R_n}{\pd\zeta} =
  -\int_0^\infty \frac{k e^{-k\zeta}}{2 + \lenBq k} J_n(ks) \dd k.
\end{equation}
For \(\lenBq = 0\), \(R'_n\) reduces to
\begin{equation}
  \label{eq:rzFn'|Bq=0}
  R_n'(0; s, |z|) = -\frac{s^n (nr + |z|)}{2 r^3 (r + |z|)^n}.
\end{equation}

To determine the flow for all $\vec x$ and $\vec y$, we can sum \cref{eq:FT-upper-fluid} and \cref{eq:FT-lower-fluid} to eliminate the Stokes double layer, which gives
\begin{equation}
  \label{eq:BIE-all-fluid}
  \FT u_i(\vec k, z; h) =
  - \frac1{\bar\mu} \FT{\mathsfi J}_{ij}(\vec k, z) \FT q_j(\vec k; h)
  + \frac1{\mu(h)} \FT{\mathsfi J}_{ij}(\vec k, z-h) F_j
\end{equation}
where
\begin{equation*}
  \FT{\vec q}(\vec k; h) = \bar\mu \lr(){%
    \frac{\FT{\vec t^1}(\vec k)}{\mu_1} - \frac{\FT{\vec t^2}(\vec k)}{\mu_2}
  }
\end{equation*}
is the Stokes single-layer density in Fourier space.
For a clean interface, setting \(z=0\) in \cref{eq:BIE-all-fluid} and putting \cref{eq:FT-vel-surf-clean} into the result yields
\begin{equation}
  \label{eq:FT-single-layer-density-clean}
  \FT{\vec q}(k;h) = 4 k \lr(){\ten I - \frac{\mu(h)}{\bar\mu} \ten I_\surf} \vdot
  \FT{\mathsfbi J}(\vec k, -h) \vdot \vec F.
\end{equation}
After inserting \cref{eq:FT-single-layer-density-clean} back into \cref{eq:BIE-all-fluid}, lengthy algebraic manipulation and inversion of $\FT{\vec u}$ to real space yields the velocity field in terms of the hydrodynamic image system \cref{eq:Gfun-clean}, with \(\vec u \equiv \ten G \vdot \vec F\).
A similar procedure may be used to fully determine \(\ten H(\vec x, \vec y)\), but we do not require that result in this article.
See also \citet{Blawzdziewicz1999} and \citet{Fischer2006} for computations of \(\ten H\) via different approaches.


\bibliographystyle{jfm}
\bibliography{main}

\end{document}